\documentclass[aps,prd,twocolumn,superscriptaddress,groupedaddress]{revtex4}  
\usepackage{graphicx}  
\usepackage{dcolumn}   
\usepackage{bm}        
\usepackage{amssymb}   

\begin{document}



\title{Study of galaxy distributions with SDSS DR14 data 
and measurement of neutrino masses}

\author{B.~Hoeneisen} \affiliation{Universidad San Francisco de Quito, Quito, Ecuador}
\date{June 4, 2018}

\begin{abstract}
\noindent
We study galaxy distributions with 
Sloan Digital Sky Survey SDSS DR14 data
and with simulations searching for variables that can
constrain neutrino masses. To be specific, we consider
the scenario of three active neutrino eigenstates with
approximately the same mass, so $\sum m_\nu = 3 m_\nu$.
Fitting the predictions of the
$\Lambda$CDM model to the Sachs-Wolfe effect, $\sigma_8$,
the galaxy power spectrum
$P_\textrm{gal}(k)$, fluctuations of galaxy counts in
spheres	of radii ranging from $16/h$ to	$128/h$	Mpc,    
BAO measurements,
and $h = 0.678 \pm 0.009$, in various combinations, \textit{with
free spectral index $n$}, and free galaxy bias \textit{and 
galaxy bias slope}, we obtain consistent          
measurements of	$\sum m_\nu$. The results depend on
$h$, so we have presented
confidence contours in the $(\sum m_\nu, h)$ plane.   
A global fit with
$h = 0.678 \pm 0.009$ obtains
$\sum m_\nu = 0.719 \pm 0.312 \textrm{ (stat)}^{+0.055}_{-0.028}
  \textrm{ (syst)}$ eV,
and the amplitude and spectral index of 
the power spectrum of linear density fluctuations $P(k)$:
$N^2 = (2.09 \pm 0.33) \times 10^{-10}$, and
$n = 1.021 \pm 0.075$.
The fit also returns the galaxy bias $b$ including
its scale dependence.
\end{abstract}

\maketitle

\section{Introduction}

We measure neutrino masses by comparing the predictions
of the $\Lambda$CDM model with measurements of the
power spectrum of linear density perturbations $P(k)$.
We consider three measurements of $P(k)$:
(i) the Sachs-Wolfe effect of fluctuations
of the Cosmic Microwave Background (CMB)
which is a direct measurement of density
fluctuations \cite{SW, Weinberg}; (ii) the
relative mass fluctuations $\sigma_8$ in randomly placed 
spheres of radius $r_s = 8/h$ Mpc with
gravitational lensing and studies of rich galaxy
clusters \cite{Weinberg, PDG2016}; and
(iii) measurements of $P(k)$
inferred from galaxy clustering with the
Sloan Digital Sky Survey \cite{Pk_BOSS, SDSS_DR14, BOSS}.
Baryon Acoustic Oscillations (BAO) were considered
separately \cite{BH_ijaa, BH_ijaa_2} and are not included in the present
study, except for the final combinations.

To be specific, we consider three active neutrino
eigenstates with nearly the same mass, so $\sum m_\nu = 3 m_\nu$.
This is a useful scenario to consider because the current
limits on $m_\nu^2$ are much larger than the
mass-squared-differences $\Delta m^2$ and $\Delta m_{21}^2$ obtained
from neutrino oscillations \cite{PDG2016}.

\begin{figure}
\includegraphics[scale=0.47]{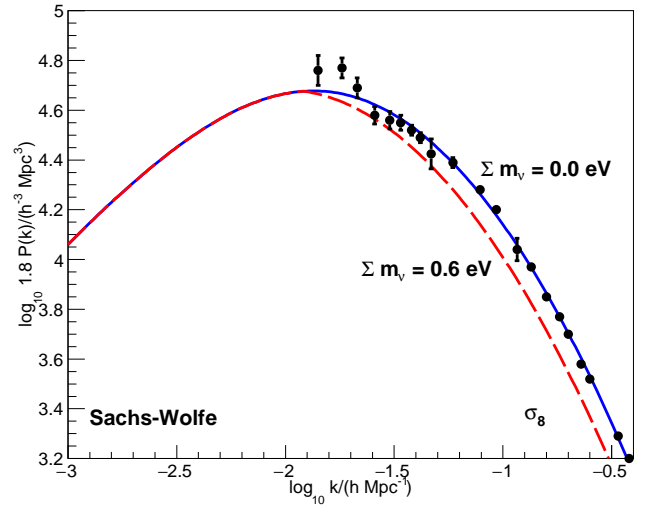}
\caption{\label{P_vs_k} 
Comparison of $P_\textrm{gal}(k)$ obtained
from the SDSS-III BOSS survey \cite{Pk_BOSS}
(``reconstructed") with $b^2 P(k)$
obtained from a 
fit of Eq. (\ref{Pk_prime}) with $\sum m_\nu = 0$ eV
to the Sachs-Wolfe effect, $\sigma_8$,
and $P_\textrm{gal}(k)$.
The fit obtains
$A = 8738 \textrm{ Mpc}^3$,
$k_\textrm{eq} = 0.068 h \textrm{ Mpc}^{-1}$,
$\eta = 4.5$, and $b^2 = 1.8$,
with $\chi^2 = 24.7$ for 19 degrees of freedom.
Also shown for comparison is the curve with
the same parameters, except $\sum m_\nu = 0.6$ eV.
}
\end{figure}

\begin{figure}
\includegraphics[scale=0.47]{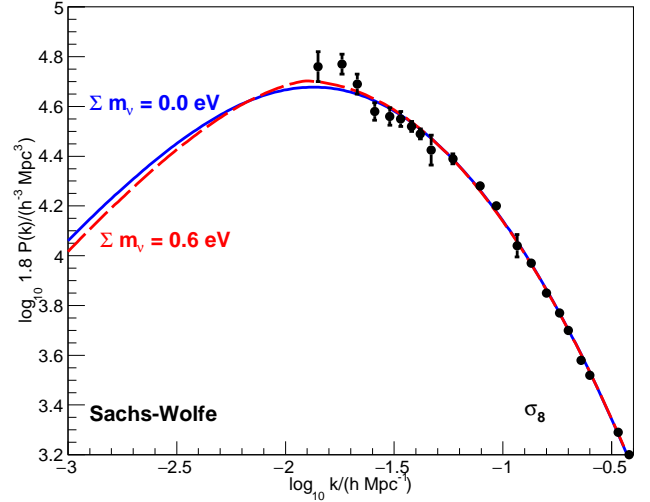}
\caption{\label{Pk_graph_log_2_wide}
Same as Fig. \ref{P_vs_k}, except that the curve
``$\sum m_\nu = 0.6$ eV" is fit.
We obtain
$A = 9312 \textrm{ Mpc}^3$,
$k_\textrm{eq} = 0.080 h \textrm{ Mpc}^{-1}$,
$\eta=4.2$, and $\kappa = 1.8$,
with $\chi^2 = 21.8$ for 19 degrees of freedom.
Note that $\sum m_\nu$ is largely degenerate with
the remaining parameters in Eq. (\ref{Pk_prime}),
unless we are able to constrain $k_\textrm{eq}$.
}
\end{figure}

\begin{figure}
\includegraphics[scale=0.47]{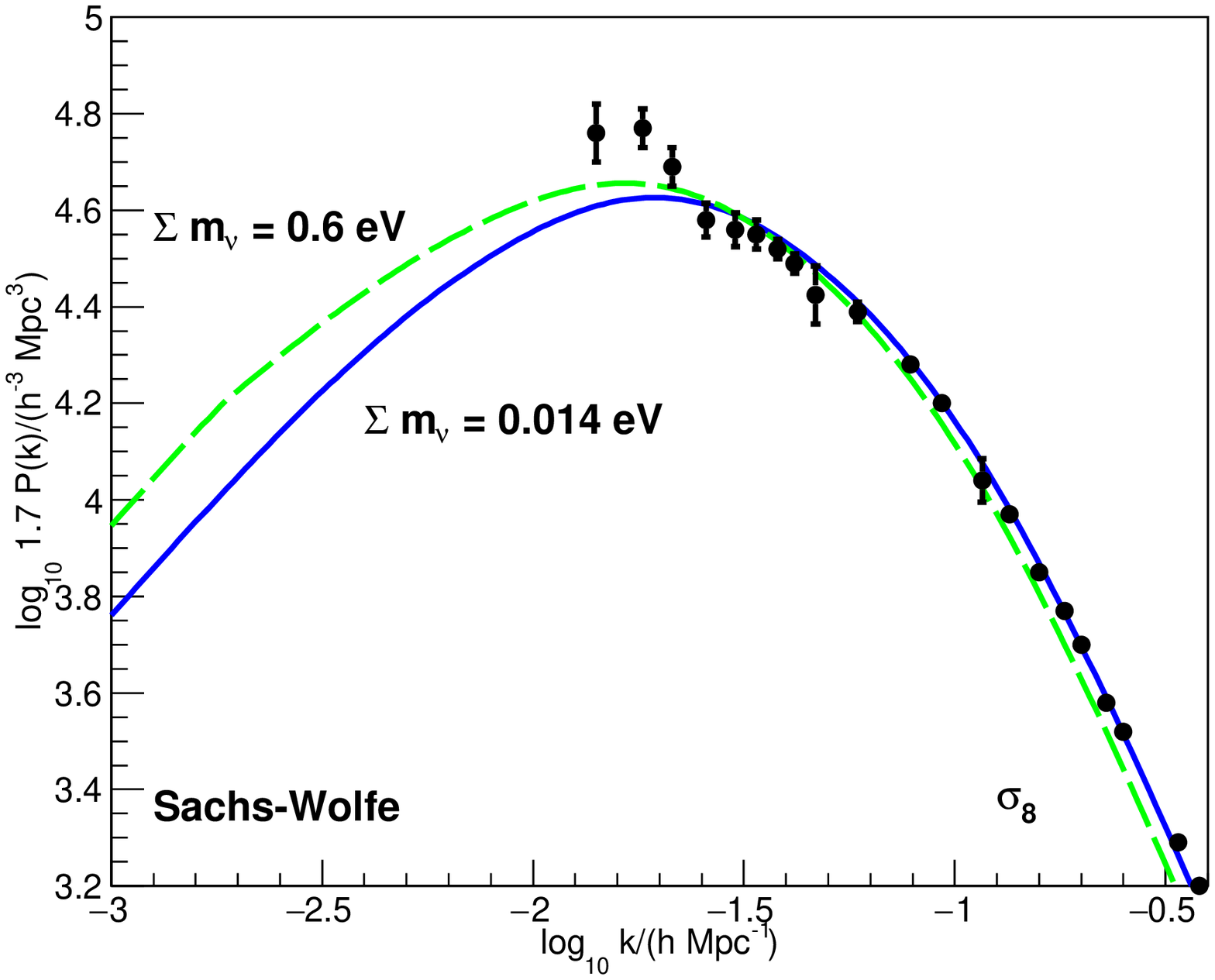}
\caption{\label{Pk_graph_log_2_wide_Weinberg}
Comparison of $P_\textrm{gal}(k)$ obtained
from the SDSS-III BOSS survey \cite{Pk_BOSS}
(``reconstructed") with $b^2 P(k)$ 
obtained from a	
fit of Eq. (\ref{Pk})
to the Sachs-Wolfe effect, $\sigma_8$,
and $P_\textrm{gal}(k)$.
The fit obtains $\sum m_\nu = 0.014 \pm 0.079$ eV, 
$N^2 = (1.41 \pm 0.12) \times 10^{-10}$, and
$b^2 = 1.7 \pm 0.1$, with $\chi^2 = 47$ for 20 degrees of
freedom (so the statistical uncertainties shown
need to be multiplied by $\sqrt{47/20}$). 
Also shown is the fit with $\sum m_\nu = 0.6$ eV fixed for
comparison.
}
\end{figure}

\begin{figure}
\includegraphics[scale=0.47]{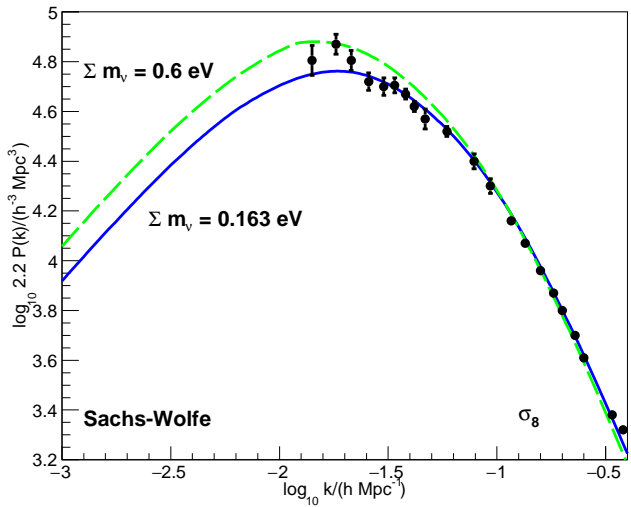}
\caption{\label{Pk_graph_log_2_wide_Weinberg_std}
Comparison of $P_\textrm{gal}(k)$ obtained
from the SDSS-III BOSS survey \cite{Pk_BOSS}
(``standard") with $b^2 P(k)$
obtained from a
fit of Eq. (\ref{Pk})                         
to the Sachs-Wolfe effect, $\sigma_8$,
and $P_\textrm{gal}(k)$.
The fit obtains $\sum m_\nu = 0.163 \pm 0.061$ eV, 
$N^2 = (1.56 \pm 0.12) \times 10^{-10}$, and
$b^2 = 2.2 \pm 0.2$, with $\chi^2 = 33.9$ for 20 degrees of
freedom (so the statistical uncertainties shown
need to be multiplied by $\sqrt{33.9/20}$). 
Also shown is the fit with $\sum m_\nu = 0.6$ eV fixed for
comparison.
}
\end{figure}

Figures \ref{P_vs_k} to \ref{Pk_graph_log_2_wide_Weinberg_std} 
illustrate the problem
at hand. Figures \ref{P_vs_k}, \ref{Pk_graph_log_2_wide}, 
and \ref{Pk_graph_log_2_wide_Weinberg}
present measurements of the ``reconstructed" 
galaxy power spectrum $P_\textrm{gal}(k)$
obtained from the SDSS-III BOSS survey \cite{Pk_BOSS},
while Fig. \ref{Pk_graph_log_2_wide_Weinberg_std} presents the
corresponding ``standard" $P_\textrm{gal}(k)$.
The ``reconstructed" $P_\textrm{gal}(k)$ is obtained from the
directly measured ``standard" $P_\textrm{gal}(k)$ by subtracting
peculiar motions to obtain the power spectrum 
prior to non-linear clustering.
Also shown are various fits to this data 
(with floating normalization), and
to measurements of 
the Sachs-Wolfe effect,
and $\sigma_8$.
The Sachs-Wolfe effect normalizes $P(k)$, within its uncertainty, in 
the approximate range of $\log_{10}( k/(h \textrm{ Mpc}^{-1})$ 
from -3.1 to -2.7,
while $\sigma_8$ is most sensitive to the range
-1.3 to -0.6. 
Full details will be given in the main body of this article.

The fit in Fig. \ref{P_vs_k} corresponds to the function
\begin{equation}
P'(k) \equiv \frac{(2 \pi)^3 a_{20}^2 A w^n}{(1 + \eta w + w^2)^2},
\label{P_prime}
\end{equation}
where $w \equiv k/k_\textrm{eq}$.
Unless otherwise noted, we take the Harrison-Zel'dovich
index $n = 1$ which is close to observations.
The parameters $A$, $\eta$, and $k_\textrm{eq}$,
as well as the normalization factor $b^2$, are free
in the fit.
The uncertainties of two data points that
fall on BAO peaks are multiplied by three (since BAO is
not included in $P'(k)$).

Also shown in Fig. \ref{P_vs_k} is the suppression
of $P(k)$ for $k$ greater than
\begin{equation}
k_\textrm{nr} = 0.018 \cdot \Omega_m^{1/2} 
\left( \frac{\sum m_\nu}{1 \textrm{eV}} \right)^{1/2} h 
\textrm{ Mpc}^{-1}
\label{knr}
\end{equation}
due to free-streaming
of massive neutrinos that can not cluster on these 
small scales, and, more importantly, to the
slower growth of structure with massive neutrinos \cite{f_mnu}.
The suppression factor for $k \gg k_\textrm{nr}$ 
for one massive neutrino, or
three degenerate massive neutrinos, is
\begin{equation}
f(k, \sum m_\nu) \equiv \frac{P(k)^{f_\nu}}{P(k)^{f_\nu = 0}} = 1 - 8 f_\nu,
\end{equation}
where $f_\nu = \Omega_\nu / \Omega_m$ \cite{f_mnu}.
$\Omega_m$ is the total (dark plus baryonic plus neutrino) 
matter density today relative to
the critical density, and includes the contribution
$\Omega_\nu$ of neutrinos that are non-relativistic today.
$\Omega_\nu = h^{-2} \sum{m_\nu}/93.04$ eV
for three left-handed plus right-handed
Majorana neutrino eigenstates, or 
three eigenstates of left-handed Dirac neutrinos plus three right-handed Dirac
anti-neutrinos, that are non-relativistic
today (right-handed Dirac neutrinos and left-handed Dirac
anti-neutrinos are assumed to not have reached thermal 
and chemical equilibrium with the Standard Model particles).
We take $f(k, \sum m_\nu) = 1$
for $k < k_\textrm{nr}/0.604$, and
\begin{equation}
f(k, \sum m_\nu) = 1 - 0.407 \frac{\sum m_\nu}{0.6 \textrm{ eV}} 
\left[ 1 - \left( \frac{k_\textrm{nr}}{0.604 k} \right)^{0.494} \right]
\label{f}
\end{equation}
for $k > k_\textrm{nr}/0.604$ and $\sum m_\nu < 1.1$ eV, 
for galaxy formation at a redshift $z = 0.5$ \cite{f_mnu}.

Figure \ref{Pk_graph_log_2_wide} is the same as Fig. \ref{P_vs_k}
except that the function
\begin{equation}
P(k) = P'(k) f(k, \sum m_\nu) 
\label{Pk_prime}
\end{equation}
with $\sum m_\nu = 0.6$ eV is fit. 
We see that the parameter $\sum m_\nu$
is largely degenerate with the parameters
$A$, $\eta$, and $k_\textrm{eq}$, so that 
only a weak sensitivity to $\sum m_\nu$ is obtained
unless we are able to constrain $k_\textrm{eq}$. 
The power spectrum $P'(k)$ of Eq. (\ref{P_prime})
neglects the growth of structure inside the horizon
while radiation dominates.

The fits in Figs. \ref{Pk_graph_log_2_wide_Weinberg} 
and \ref{Pk_graph_log_2_wide_Weinberg_std}
make full use of the $\Lambda$CDM theory.
The fitted function is
\begin{equation}
P(k) = P''(k) f(k, \sum m_\nu), 
\label{Pk}
\end{equation}
where
$P''(k)$ is given by \cite{Weinberg}:
\begin{equation}
P''(k) = \frac{4 (2 \pi)^3 N^2 C^2 k \tau^2(\sqrt{2}k/k_{\textrm{eq}})}{25 \Omega_m^2 H_0^4}
\left( \frac{k_\textrm{\tiny{SW}}}{k} \right)^{1-n},
\label{Weinberg_P_k}
\end{equation}
with
\begin{equation}
k_{\textrm{eq}} = \frac{\sqrt{2} H_0 ( \Omega_m - \Omega_\nu )}{\sqrt{\Omega_r}}.
\end{equation}
$C$ is a function of $\Omega_\Lambda / \Omega_m$, and we
take $C = 0.767$ \cite{Weinberg}. $\tau(\sqrt{2}k/k_{\textrm{eq}})$ 
is a function given in Reference \cite{Weinberg}.
The pivot point
$k_\textrm{\tiny{SW}} = 0.001 \textrm{ Mpc}^{-1}$ is
chosen to not upset Eq. (\ref{Nsq}) below.
The fit depends on $h$, $\Omega_m$, and the spectral index $n$,
so we define $\delta h \equiv (h - 0.678)/0.009$ \cite{PDG2016},
$\delta \Omega_m \equiv (\Omega_m - 0.2810)/0.003$ \cite{BH_ijaa}, and
$\delta n \equiv (n - 1)/0.038$ \cite{PDG2016},
and obtain, tentatively,
\begin{eqnarray}
\sum{m_\nu} & = & 0.014 + 0.162 \cdot \delta h +
0.807 \cdot \delta n + 0.142 \cdot \delta \Omega_m \nonumber \\
& & \pm 0.079 \sqrt{47/20} \textrm{ (stat)} ^{+0.005}_{-0.007} \textrm{ (syst) eV},
\label{mnu1}
\end{eqnarray}
by minimizing the $\chi^2$ with respect to
$\sum m_\nu$, and $N^2$.
The statistical uncertainty has been multiplied by
the square root of the $\chi^2$ per degree of freedom.
This result corresponds to the ``reconstructed" data in 
Fig. \ref{Pk_graph_log_2_wide_Weinberg}.
The systematic uncertainties included are from the top-hat
window function instead of the gaussian window function,
and an alternative value of $\sigma_8$
(details will be given in Section \ref{section_sigma8}).
\textit{Not included} is the systematic uncertainty due to the
possible scale dependence of the galaxy bias $b$.

To obtain $P(k)$, 
we would like to measure the density $\rho(\vec{r}, z)$
at redshift $z$, but we only have information
on the peaks of $\rho(\vec{r}, z)$ that have gone
non-linear collapsing into visible galaxies.
How accurate is the measurement of $P(k)$ with galaxies?
The measurement of $P_\textrm{gal}(k)$ in Ref. \cite{Pk_BOSS} is based
on a procedure described in \cite{Feldman} based on
``the usual assumption that the galaxies form a
Poisson sample \cite{Peebles} of the density field".
In other words, the assumption is that the number density
of point galaxies $n(x)$ is equal to its expected mean 
$\bar{n}$ (which depends on the position dependent
galaxy selection criteria),
modulated by the perturbation
of the density field:
\begin{equation}
\frac{n(\vec{x})}{\bar{n}} = b \frac {\rho(\vec{x})}{\bar{\rho}} 
\equiv b (1 + \delta_c(\vec{x})).
\label{n_x}
\end{equation}
Both sides of this equation are measured or calculated
at the same length scale, and at the same time.
The ``galaxy bias" $b$ is explicitly assumed to be
scale invariant.
If we choose a region of space such that $\bar{n}$ is constant,
then the galaxy power spectrum $P_\textrm{gal}(k)$
(derived from $n(\vec{x})/\bar{n}$)
should be proportional, under the above assumption, 
to the power spectrum of linear density
perturbations $P(k)$ (derived from $\delta_c(\vec{x})$)
up to corrections: 
\begin{equation}
P_\textrm{gal}(k) = b^2 P(k).
\label{Pgal_Pb}
\end{equation}
It is due to this bias $b$ that we have freed the normalization of
the measured $P_\textrm{gal}(k)$ in the fits corresponding to Figs. \ref{P_vs_k} to 
\ref{Pk_graph_log_2_wide_Weinberg_std}.

In the following Sections we study galaxy distributions
with SDSS DR14 data and with simulations, in order to understand
their connection with the underlying power spectrum
of linear density fluctuations $P(k)$. In the end we return
to the measurement of neutrino masses.

\section{The hierarchical formation of galaxies}
\label{simulation}

\begin{figure}
\includegraphics[scale=0.47]{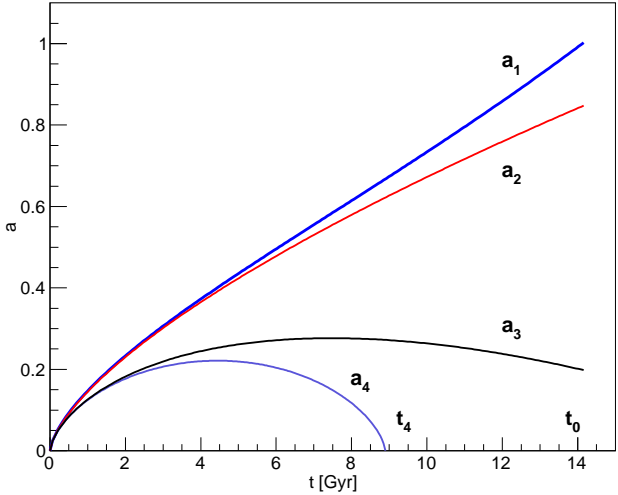}
\caption{\label{a_vs_t} 
Expansion parameters as a function of time $t$
of four solutions of the Friedmann equation. 
From top to bottom, $a_1$
corresponds to 
$(\Omega_m, \Omega_\Lambda, \Omega_k) = (0.281, 0.719, 0)$,
$a_2$ corresponds to $(0.281, 0, 0)$,
$a_3$ is the linear approximation to $(0.281, 0, -1.27)$,
while $a_4$ is the exact solution to $(0.281, 0, -1.27)$.
$a_4$ is the exact solution for the growing mode of
a spherically symmetric density peak that collapses
to a galaxy at $t_4$.
In all cases $h = 0.678$.
}
\end{figure}

This Section allows a precise definition of $P(k)$,
and an understanding of the connection between $P(k)$ and $P_\textrm{gal}(k)$.
We generate galaxies as follows (see \cite{BH_galaxies} for full details).
The evolution of the Universe in the
homogeneous approximation is described by the Friedmann equation
\begin{equation}
\frac{1}{H_0} \frac{1}{a_1} \frac{d a_1}{d t} \equiv E(a_1) =
\sqrt{\frac{\Omega_r}{a_1^4} + \frac{\Omega_m}{a_1^3} + 
\frac{\Omega_k}{a_1^2} + \Omega_\Lambda}.
\label{Friedmann}
\end{equation}
The expansion parameter $a_1(t)$ has been normalized to 1 at the present
time $t_0$: $a_1(t_0) = 1$. $H_0$ has been normalized so $E(1) = 1$.
Therefore $H_0$ is the present Hubble expansion rate. With these normalizations
we have $\Omega_r + \Omega_m + \Omega_k + \Omega_\Lambda \equiv 1$. 
The matter density
is $\rho_m(t) = \Omega_m \rho_c /a_1^3$, where
$\rho_c = 3 H_0^2/(8 \pi G_N)$ is the critical density of the Universe.
We are interested in the period after the density of matter
exceeds the density of radiation.
For our simulations we assume flat space, i.e. $\Omega_k = 0$,
we neglect the radiation density $\Omega_r$, 
take $\Omega_\Lambda = 0.719$ constant \cite{BH_ijaa},
and the present Hubble expansion rate $H_0 = 100 h$ km s$^{-1}$ Mpc$^{-1}$ with
$h = 0.678$ \cite{PDG2016}.
The solution to Eq. (\ref{Friedmann}) with these parameters is shown
by the curve ``$a_1$" in Fig. \ref{a_vs_t}. The present age of the 
universe with these parameters is $t_0 = 14.1$ Gyr.

Setting $\Omega_\Lambda = 0$ we obtain the critical universe
with expansion parameter 
\begin{equation}
a_2 = \left[ \frac{3 H_0}{2} \right]^{2/3} \Omega_m^{1/3} t^{2/3},
\end{equation}
also shown in Fig. \ref{a_vs_t}. We note that $a_2(t_0) \equiv a_{20}= 0.846$.
Let us now add density fluctuations to this critical universe
and consider a density peak. The growing mode for this 
density peak is obtained by adding a negative $\Omega_k$ to
the critical Universe. 
This prescription is exact if the density peak is spherically
symmetric.
An example with ``expansion parameter"
$a_4$ is presented in Fig. \ref{a_vs_t}. Note that
$a_4$ grows to maximum expansion and then collapses
to zero at time $t_4 = \pi \Omega_m / [ (-\Omega_k)^{3/2} H_0]$, 
and, in our model \cite{BH_galaxies},
a galaxy forms. In the example of Fig. \ref{a_vs_t} the galaxy
forms at redshift $z = 0.5$.
$a_3(t)$ is the linear approximation to $a_4(t)$.
In the linear approximation for growing modes the density
fluctuations relative to $\rho_2$ grow in proportion to $a_2(t)$:
\begin{equation}
\delta_c \equiv \frac{\rho_3 - \rho_2}{\rho_2}
= 3 \frac{a_2 - a_3}{a_2} = -\frac{3 \Omega_k}{5 \Omega_m} a_2(t),
\end{equation}
while $\delta_c \ll1$.
At the time $t_4$, when the galaxy forms,
$\delta_c \equiv (\rho_3 - \rho_2) / \rho_2 = 1.69$
in the linear approximation (which has already broken down).

In the linear approximation
the density due to Fourier components of wavevector $| \vec{k} | \le k_I$ is
\begin{equation}
\rho_\textrm{lin}(\vec{x}, t, I) = \frac{\Omega_m \rho_c}{a_2^3}
\left\{ 1 + \delta_c(\vec{x}, t, I) \right\},
\end{equation}
where
\begin{equation}
\delta_c(\vec{x}, t, I) = 
\sum_{\vec{k}, k \le k_I} | \delta_{\vec{k}} | a_2(t) 
\exp{ \left[ i \frac{ \vec{k} \cdot \vec{x} }{a_2(t)} + i \varphi_{\vec{k}} \right] }.
\label{delta_c}
\end{equation}
$\varphi_{\vec{k}}$ are random phases. 
The sum of the Fourier series is over comoving wavevectors
that satisfy periodic boundary conditions in a rectangular
box of volume $V = L_x L_y L_z$:
\begin{equation}
\vec{k} = 2 \pi \left( \frac{n}{L_x} \hat{i} + \frac{m}{L_y} \hat{j} 
+ \frac{l}{L_z} \hat{k} \right),
\label{k}
\end{equation}
where $L_x = n_\textrm{max} L_0$, $L_y = m_\textrm{max} L_0$, $L_z = l_\textrm{max} L_0$,
$n, m, l = 0, \pm 1, \pm 2, \pm 3, \cdot \cdot \cdot$, and
\begin{equation}
k_I = k_\textrm{max} \frac{I}{I_\textrm{max}}
\end{equation}
where $k_\textrm{max} = 2 \pi / L_0$, and
$I = 1, 2, ... I_\textrm{max}$.

Inverting Eq. (\ref{delta_c}) obtains
\begin{eqnarray}
\delta_{\vec{k}} & \equiv & | \delta_{\vec{k}} | e^{i \varphi_{\vec{k}}}
= \delta^{*}_{-\vec{k}} \nonumber \\
& = & \frac{1}{V} \int_V \frac{\delta_c(\vec{x}, t)}{a_2(t)}
\exp{ \left( -i \vec{k} \cdot \vec{X} \right) } d^3\vec{X},
\label{d_k}
\end{eqnarray}
where $\vec{X} \equiv \vec{x}(t)/a_2(t)$ is the comoving coordinate in
the linear approximation.
The power spectrum of density fluctuations 
\begin{equation}
P(\vec{k}) \equiv V | \delta_{\vec{k}} |^2 a^2_{20}
\label{P_k}
\end{equation}
is defined in the linear approximation corresponding
to $a_3$, and is approximately independent of $V$ for large $V$. 
Averaging over $\vec{k}$ in a bin of $k \equiv |\vec{k}|$ obtains $P(k)$.
Note that
\begin{equation}
\frac{1}{V}
\sum_{\vec{k}} P( \vec{k} ) = a^2_{20} \sum_{\vec{k}} | \delta_{\vec{k}} |^2
= \frac{1}{V} \int_V \delta^2_c(\vec{x}, t_0) d^3 \vec{X}.
\end{equation}
Each term in this equation is approximately independent of $V$.
The Fourier transform of the power spectrum is the correlation function:
\begin{equation}
\sum_{\vec{k}} P( \vec{k} ) e^{i \vec{k} \cdot \vec{X} } = 
\int_V \delta_c(\vec{x}', t_0) \delta_c(\vec{x}' + \vec{x}, t_0) d^3 \vec{X'}.
\end{equation}

\begin{figure}
\includegraphics[scale=0.47]{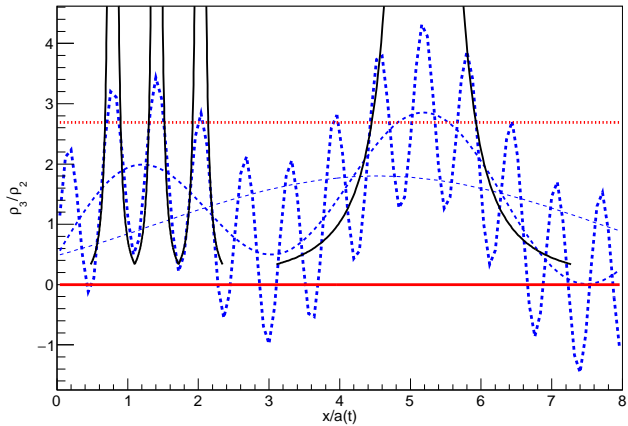}
\caption{\label{hierarchical_galaxies} 
The hierarchical formation of galaxies \cite{BH_galaxies}.
Three Fourier components of the density 
in the linear approximation are shown.
Note that in the linear approximation
$\delta_c \equiv (\delta_3 - \delta_2)/\delta_2 \propto a(t)$.
When $\delta_c$ reaches 1.69 in the
linear approximation the exact solution diverges
and a galaxy forms.
As time goes on, density perturbations grow, 
and groups of galaxies of one generation coalesce into larger
galaxies of a new generation as shown on the right.
}
\end{figure}

The generation of galaxies at time $t$ proceeds as follows. We 
start with $I = 2$, calculate $\delta_c(\vec{x}, t, I)$, and search
for local maximums of $\delta_c(\vec{x}, t, I)$ inside a 
comoving volume $L_x L_y L_z$. If a maximum
exceeds 1.69 we generate a galaxy 
of radius 
\begin{equation}
R(I) \approx \frac{\pi a_2}{k_I}, 
\end{equation}
and dark plus baryonic plus neutrino mass 
\begin{equation}
M(I) \approx 2.69 \frac{4 \pi R(I)^3}{3} \frac{\Omega_m \rho_c}{a_2^3}
\label{M}
\end{equation}
if it ``fits", i.e. if it
does not overlap previously generated galaxies. $I$ is 
increased by 1 unit to generate galaxies of a smaller
generation, until $I = I_\textrm{max}$ is reached.
See Fig. \ref{hierarchical_galaxies}.

The peculiar velocity of the generated galaxies is
\begin{equation}
\vec{v}_\textrm{pec}(\vec{x}, t) = \sum_{\vec{k}, k < k_{I - 1}} 
\frac{2 i {\vec{k}} a_2^2 | \delta_{\vec{k}} |}{3 k^2 t}
\exp{ \left[ i \frac{ \vec{k} \cdot \vec{x} }{a_2} + i \varphi_{\vec{k}} \right] },
\end{equation}
and their peculiar displacement is
\begin{equation}
\vec{x}_{\textrm{pec}} = \frac{3}{2} \vec{v}_{\textrm{pec}} t.
\end{equation}
$\vec{x} + \vec{x}_{\textrm{pec}}$ is the proper coordinate of a galaxy
at the time $t$ of its generation, and $a_2 \equiv a_2(t)$.
The comoving coordinate of this galaxy, i.e.
its position extrapolated to the present time, is
the corresponding $(\vec{x} + \vec{x}_{\textrm{pec}}) / a_1(t)$.
$\vec{x}_{\textrm{pec}}$ causes the difference between
the data points $P_\textrm{gal}(k)$ in Figs. \ref{Pk_graph_log_2_wide_Weinberg} 
and \ref{Pk_graph_log_2_wide_Weinberg_std} at large $k$.

Note in Fig. \ref{hierarchical_galaxies} that the formation
of galaxies is hierarchical: small galaxies
form first, and, as time goes on, density perturbations
grow, and groups of galaxies
coalesce into
larger galaxies in an ongoing process until
dark energy dominates and the hierarchical formation
of galaxies comes to an end. 
The distribution of galaxies of
generation $I$ depend only on $P(k)$ for
$k < 2 \pi I / L_0$. Also, luminous galaxies
occupy a total volume (luminous plus dark) less
than $1/2.69$ of space. 

\begin{figure}
\includegraphics[scale=0.47]{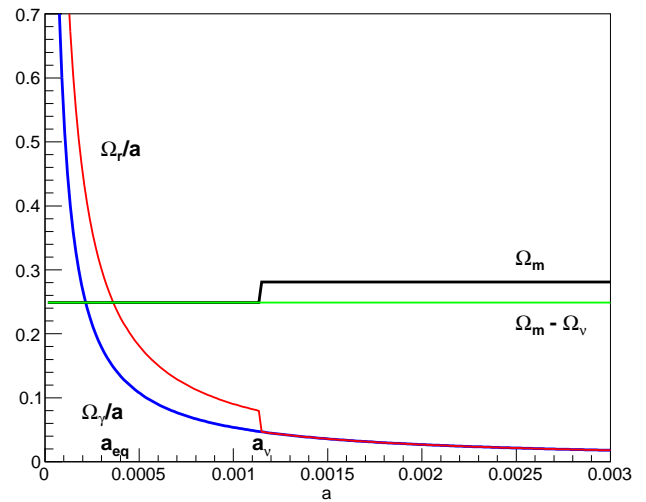}
\caption{\label{a_mnu}
Example with $m_\nu = 0.46$ eV for each of 3 active
neutrino eigenstates. 
Neutrinos become non-relativistic at $a_\nu \approx 0.00115$.
The matter density relative to the critical density
is $\Omega_m - \Omega_\nu$ for $a < a_\nu$, and
$\Omega_m$ for $a > a_\nu$. The densities of matter
and radiation become equal at $a_{\textrm{eq}} = 0.00036$.
}
\end{figure}

Neutrinos with $0 < m_\nu < 1.17$ eV become
non-relativistic after the densities of radiation and
matter become equal, as illustrated in Fig. \ref{a_mnu}.

\section{Fluctuation amplitude $\sigma_8$}
\label{section_sigma8}

$\sigma_8$ is the root-mean-square fluctuation of total mass
relative to the mean
in randomly placed volumes of radius $r_s = 8 h^{-1}$ Mpc.
We use a ``gaussian window function"
\begin{equation}
W(r) = \frac{1}{V_W} \exp{ \left( - \frac{r^2}{2r_W^2} \right) },
\end{equation}
which smoothly defines a volume
\begin{equation}
V_W \equiv \frac{4}{3} \pi r_s^3 = (2 \pi)^{3/2} r_W^3.
\end{equation}
Note that
\begin{equation}
\int_0^{\infty} W(r) 4 \pi r^2 dr = 1.
\end{equation}
The Fourier transform of $W(r)$ is
\begin{equation}
W(k) \equiv \int W(r) e^{-i \vec{k} \cdot \vec{r}} d^3 r = \exp{(-k^2 r_W^2 / 2)}. 
\label{W_k}
\end{equation}
Then
\begin{equation}
\sigma_8^2 = \frac{1}{(2 \pi)^3}
\int_0^{\infty} { 4 \pi k^2 dk P(k) \exp{ \left( -k^2 r_W^2 \right) } }.
\label{s8}
\end{equation}
An alternative window function is the ``top hat" function
$f(r) = 3/(4 \pi r_s^3)$ for $r < r_s$, and $f(r) = 0$ for $r > r_s$.
Then 
\begin{equation}
f(k) = \frac{3}{(k r_s)^3} \left( \sin{(k r_s)} - k r_s \cos{(k r_s)} \right).
\end{equation}
Direct measurements obtain \cite{PDG2016}
\begin{equation}
\sigma_8 = \left[ 0.813 \pm 0.013 \textrm{ (stat)} \pm 0.024 \textrm{ (syst)} \right]
(\Omega_m/0.25)^{-0.47}.
\label{sigma_8}
\end{equation}
80\% of $\sigma_8^2$ is due to $k/h$ in the range 0.05 to 0.25 Mpc$^{-1}$.
For comparison, from the 6-parameter $\Lambda$CDM fit \cite{PDG2016},
$\sigma_8 = 0.815 \pm 0.009$.

\section{The Sachs-Wolfe effect}

The spherical harmonic expansion of the CMB temperature fluctuation is
\begin{equation}
\Delta T(\theta, \phi) \equiv T(\theta, \phi) - T_0 =
\sum_{lm} a_{lm} Y_{lm}(\theta, \phi).
\end{equation}
Averaging over $m$ obtains $C_l \equiv \left< |a_{lm}|^2 \right>$.
The variable that is measured is \cite{Weinberg}
\begin{equation}
\left< \Delta T(\hat{n}_1) \Delta T(\hat{n}_2) \right> =
\sum_l{ C_l \frac{2l + 1}{4 \pi}} P_l(\hat{n}_1 \cdot \hat{n}_2).
\end{equation}
For $7 < l < 20$ the dominant contribution to $C_l$ is from 
the Sachs-Wolfe effect \cite{SW, Weinberg, PDG2016}. This range 
corresponds to $0.0007 \text{ Mpc}^{-1} < k/h < 0.002 \text{ Mpc}^{-1}$.
The Sachs-Wolfe effect relates temperature fluctuations
of the CMB
to perturbations of the gravitational potential $\phi$ \cite{Weinberg}:
\begin{equation}
\left( \frac{\Delta T(\hat{n})}{T_0} \right)_{\textrm{SW}} 
= \frac{1}{3} \delta \phi(\hat{n}).
\end{equation} 
When expressed as a function of comoving coordinates, $\phi(\vec{X})$
is independent of time when matter dominates.
The primordial power spectrum of gravitational potential 
fluctuations is assumed to have
the form \cite{Weinberg}
\begin{equation}
P_\phi(k) = N_\phi^2 k^{n-4}.
\end{equation}
The relation between $N_\phi^2$ and $N^2$ is 
$N_\phi^2 = 9 N^2 / 25$ \cite{Weinberg}. 
In the present analysis, unless otherwise stated, we assume the
Harrison-Zel'dovich power spectrum with $n = 1$, which is close to observations
\cite{PDG2016}.
For $7 \lesssim l \lesssim 20$,
\cite{Weinberg}
\begin{equation}
C_l = \frac{24 \pi Q^2}{5 l(l + 1)},
\end{equation}
where the ``quadrupole moment" $Q$ is measured to be
\begin{equation}
Q = 18.0 \pm 1.4 \mu \textrm{K}
\end{equation}
from the 1996 COBE results (see list of 
references in \cite{Weinberg}). Then, for $P'(k)$,
\begin{equation}
\frac{A}{k_\textrm{eq}} = \frac{1}{\Omega_m^2} 
\left( \frac{c}{H_0} \right)^4 \frac{12 Q^2}{5 \pi T_0^2}
= (16 \pm 3) \times 10^{4} \textrm{ Mpc}^4, 
\end{equation}
and for $P''(k)$,
\begin{equation}
N^2 = \frac{15}{\pi} \frac{Q^2}{T_0^2} = (2.08 \pm 0.33) \times 10^{-10},
\label{Nsq}
\end{equation}
independently of $\sum m_\nu$.
Detailed integration obtains results within 10\% for $5 < l < 18$.

\section{Data and simulations}
\label{Data}

\begin{figure}
\includegraphics[scale=0.47]{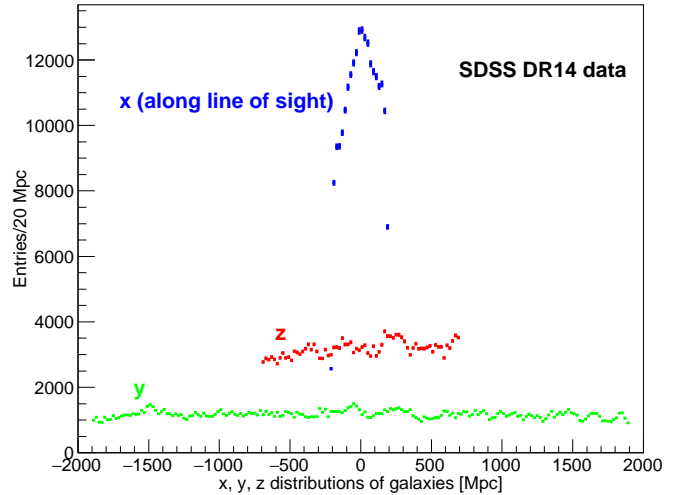}
\caption{\label{x_y_z}
Distributions of 222470 SDSS DR14 galaxies in a
``rectangular" box of dimensions
$L_x = 400$ Mpc along the line of sight 
(corresponding to redshift $z = 0.5 \pm 0.046$),
$L_y = 3800$ Mpc (corresponding to an angle $86^0$
across the sky), and
$L_z = 1400$ Mpc (corresponding to an angle $32^0$).
}
\end{figure}

The data are obtained from the publicly available
SDSS DR14 catalog \cite{SDSS_DR14, BOSS}, 
see acknowledgement. We consider objects
classified as \textsf{GALAXY}, with
redshift $z$ in the range 0.4 to 0.6,
with redshift error $\textsf{zErr} < 0.002$,
passing quality selection flags.
We further select galaxies in the northern galactic
cap, in a ``rectangular" volume
with $L_x = 400$ Mpc along the line of sight
(corresponding to redshift $z \approx 0.5 \pm 0.046$),
$L_y = 3800$ Mpc 
(corresponding to an angle $86^0$ across the sky),
and $L_z = 1400$ Mpc
(corresponding to an angle $32^0$).
In total 222470 galaxies pass these selections.
The distributions of these galaxies are shown
in Fig. \ref{x_y_z}.

Unless otherwise specified, the simulations have
$L_x = L_y = L_z = 700$ Mpc, $I_{\textrm{max}} = 59$,
$h = 0.678$, $\Omega_\Lambda = 0.719$, $\Omega_m = 0.281$,
$\Omega_k = 0$, and the input power spectrum of density 
fluctuations is (\ref{Pk_prime}) with
$A = 9200$ Mpc$^3$, $k_{\textrm{eq}}/h = 0.067$ Mpc$^{-1}$, 
$\eta = 4.7$, and $\sum m_\nu = 0$ eV.
We generate galaxies at redshift $z = a_1^{-1} - 1 = 0.5$,
corresponding to $t=8.9$ Gyr, and $a_2(t) = 0.62$.
This reference simulation has 34444 galaxies, which
is near the limit we can generate with
available computing resources.

Some definitions are in order.
For data we define the absolute red magnitude of a galaxy
MAGr at redshift $z$ as the SDSS DR14 variable \textsf{-modelMag\_r}
corrected to the reference redshift $0.35$.
Similarly, we define the absolute green magnitude of a galaxy
MAGg at redshift $z$ as the SDSS DR14 variable \textsf{-modelMag\_g}
corrected to the reference redshift $0.35$.
For a simulated galaxy we define the absolute magnitude
MAG $\equiv -19 + 2.5 \log_{10}(M/10^{16} M_\odot)$,
where $M$ is defined by Eq. (\ref{M}). 
Note that MAGr and MAGg are derived from observed luminosities,
while MAG is derived from the total (baryonic plus dark plus neutrino)
mass of the simulation. These quantities can only be compared
if the luminosity-to-mass ratio is known.

The number of galaxies per unit volume depends on the
limiting magnitude of the survey, or on $I_\textrm{max}$ of
the simulation.

\section{Distributions of galaxies in SDSS DR14
data and in simulation}

\begin{figure}
\includegraphics[scale=0.47]{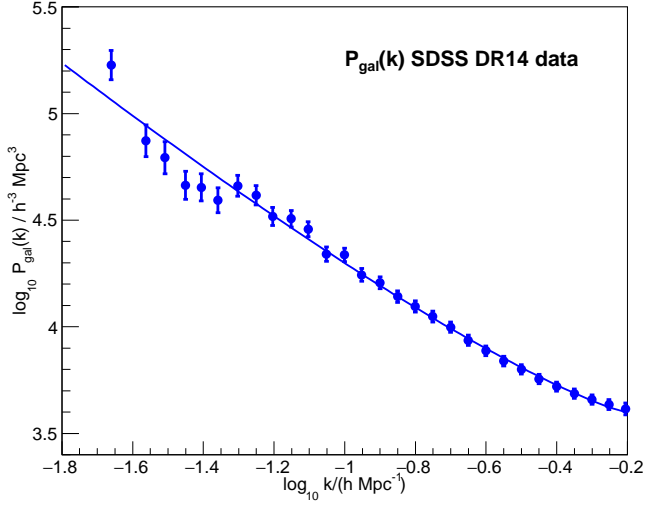}
\caption{\label{Pgal_k_SDSS_DR14}
Galaxy power spectrum (plus noise visible at large $k$) from SDSS DR14 data
in a volume $V = 400 \times 3800 \times 1400 \textrm{ Mpc}^3$,
at redshift $z = 0.5 \pm 0.0457$.
The fit is
$y = 3.60 + 0.92 \cdot (0.2 - x)^{1.23}$
with $\chi^2 = 26.8$ for 29 degrees of freedom,
where $x$ and $y$ are the axis in this figure.
}
\end{figure}

\begin{figure}
\includegraphics[scale=0.47]{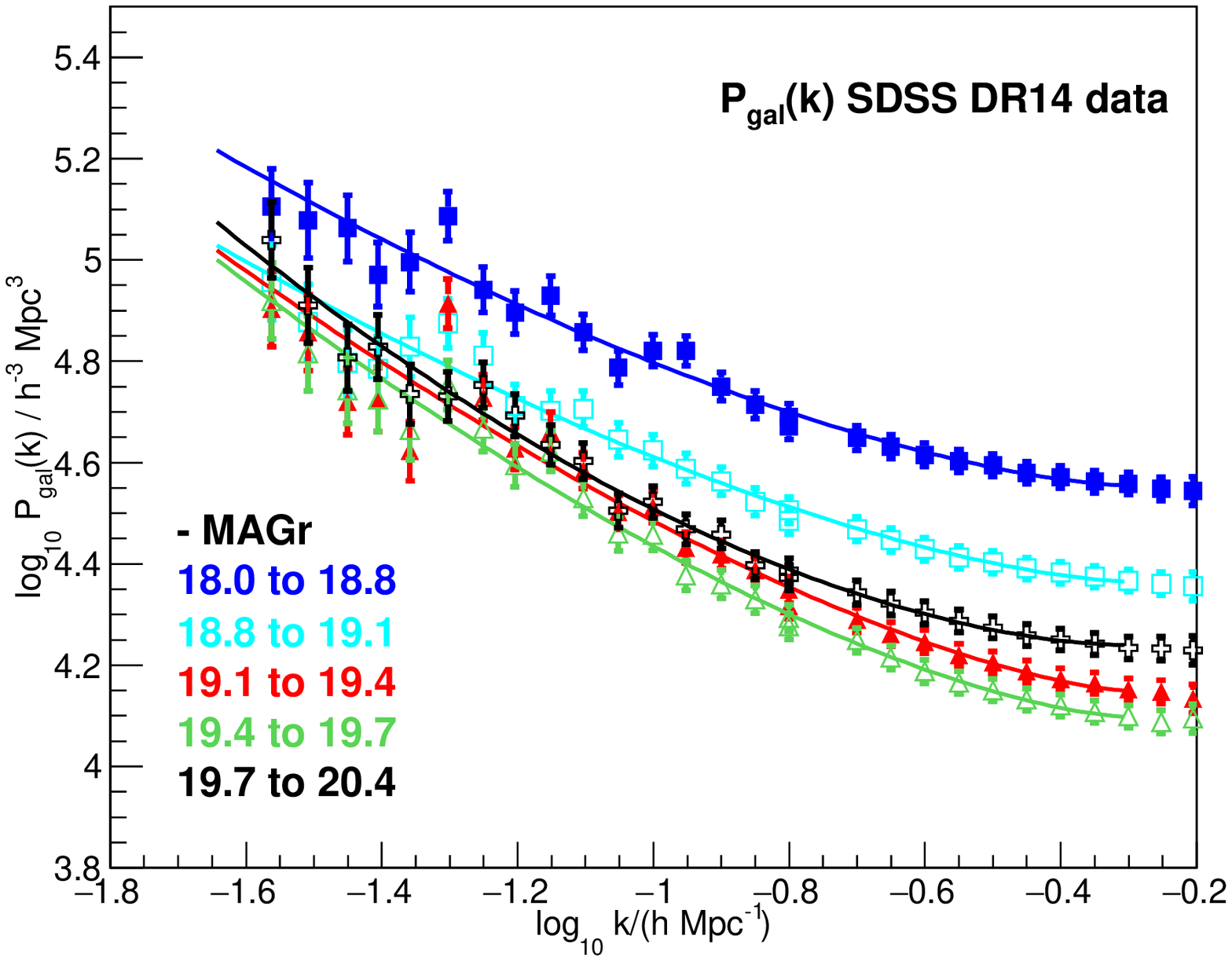}
\caption{\label{P_gal_k_SDSS14_2}
Galaxy power spectrum (plus noise visible at large $k$) in bins of MAGr from
SDSS DR14 galaxies with redshift $z = 0.5 \pm 0.0457$.
}
\end{figure}

\begin{figure}
\includegraphics[scale=0.47]{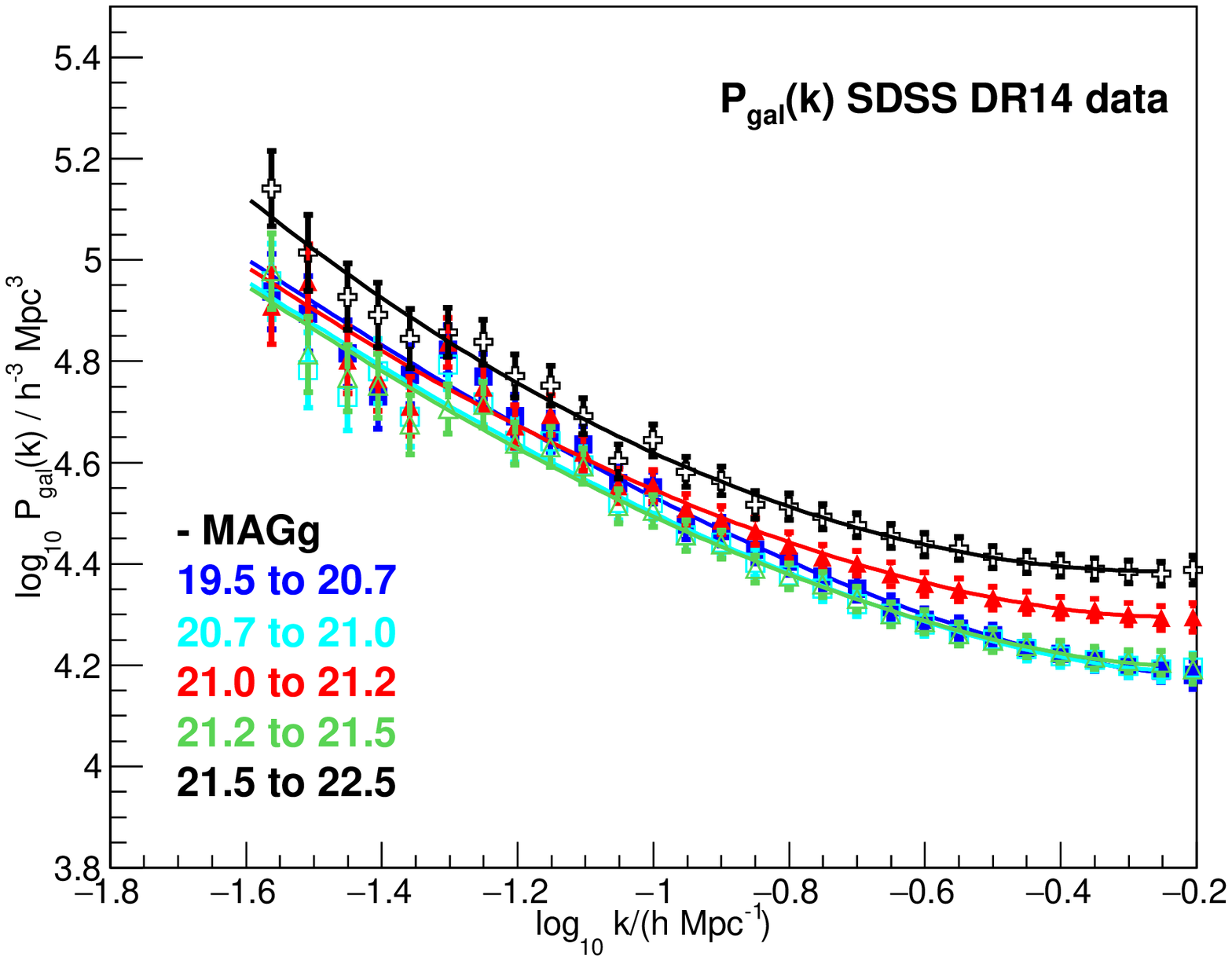}
\caption{\label{P_gal_k_SDSS14_2_g}
Galaxy power spectrum (plus noise visible at large $k$) in bins of MAGg from
SDSS DR14 galaxies with redshift $z = 0.5 \pm 0.0457$.
}
\end{figure}

\begin{figure}
\includegraphics[scale=0.47]{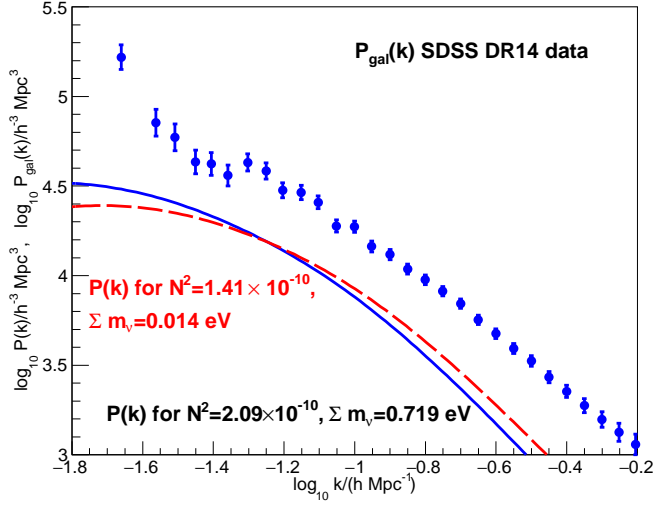}
\caption{\label{Pgal_k_SDSS_DR14_less_noise}
Noise subtracted galaxy power spectrum 
$P_\textrm{gal}(k)$, obtained from 
Fig. \ref{Pgal_k_SDSS_DR14}, compared with $P(k)$
calculated with the indicated parameters.
Their ratio is $b^2$.
}
\end{figure}

\begin{figure}
\includegraphics[scale=0.47]{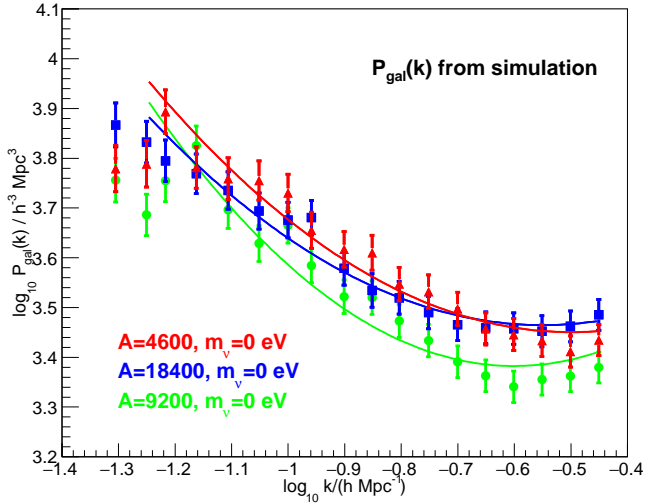}
\caption{\label{P_gal_k_MC_2_run29_run30_run31}
Galaxy power spectrum (plus noise visible at large $k$)
from simulations with three amplitudes $A$.
All other parameters of the simulation are given in
Section \ref{Data}.
}
\end{figure}

\begin{figure}
\includegraphics[scale=0.47]{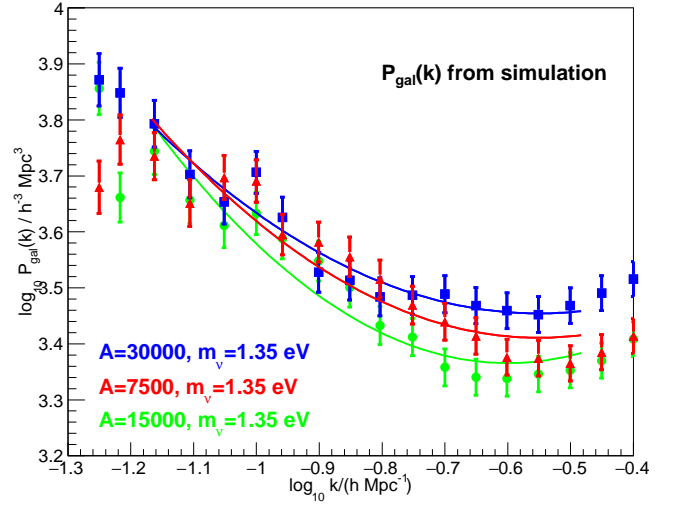}
\caption{\label{P_gal_k_MC_2_run30_run32_run34}
Galaxy power spectrum (plus noise visible at large $k$)
from simulations with $\sum m_\nu = 1.35$ eV,
with three amplitudes $A$. Other parameters are
$\eta = 4.4$, and $k_\textrm{eq}/h = 0.14 \textrm{ Mpc}^{-1}$.
}
\end{figure}

\begin{figure}
\includegraphics[scale=0.47]{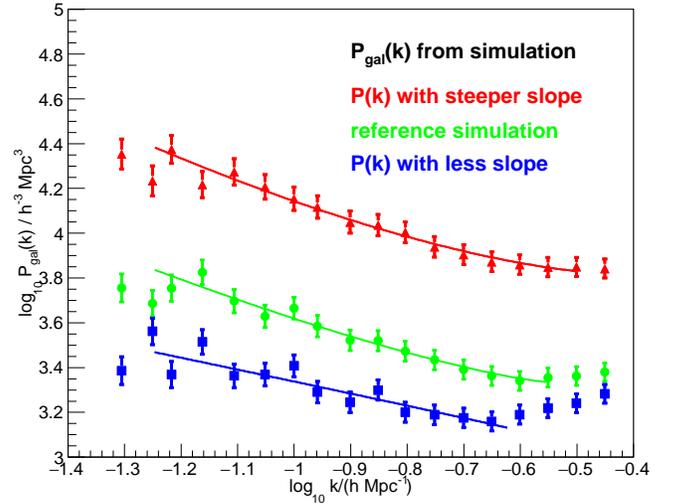}
\caption{\label{P_gal_k_MC_2_run29_run52_run53}
Galaxy power spectrum (plus noise visible at large $k$)
of the reference simulation, a simulation with
$P(k)$ with steeper slope ($P(k)$ of Eq. (\ref{P_prime}) is
multiplied by $-\log_{10} k/(h \textrm{ Mpc}^{-1})$),
and a simulation with less slope
($P(k)$ of Eq. (\ref{P_prime}) is
divided by $-\log_{10} k/(h \textrm{ Mpc}^{-1})$).
}
\end{figure}

We would like to obtain $P(k)$ from Eqs. (\ref{d_k}) and (\ref{P_k}).
Unfortunately we do not have access to the relative 
density fluctuation $\delta_c(\vec{x}, t)$. Instead we have
access to the positions of galaxies and their luminosities.
The relation between luminosity and mass of galaxies depends
on many variables and is largely unknown, so we focus on
the information contained in the positions of galaxies.

Let
\begin{equation}
n(\vec{X}) = \bar{n} \left( 1 + 
  \sum_{\vec{k}'} \Delta_{\vec{k}'} e^{i\vec{k}' \cdot \vec{X}} \right)
\label{n}
\end{equation}
be the number density of point galaxies at redshift $z$ as a function
of the comoving coordinate $\vec{X} \equiv \vec{x}(t)/a(t)$.
We have applied periodic boundary conditions in a comoving volume
$V = L_x L_y L_z$, so $\vec{k}'$ has the discrete values of Eq. (\ref{k}).
$n(\vec{X})$ is real, so $\Delta_{-\vec{k}'} = \Delta^*_{\vec{k}'}$.
The number of galaxies in $V$ is $N_{\textrm{gal}} = \bar{n} V$.
To invert Eq. (\ref{n}), we multiply it by $\exp{(-i \vec{k} \cdot \vec{X})}$,
integrate over $V$, and obtain a sum over galaxies $j$:
\begin{equation}
\sum_j e^{-i\vec{k} \cdot \vec{X_j}} = N_{\textrm{gal}} \Delta_{\vec{k}} 
  + N_{\textrm{gal}}^{1/2} e^{i \phi}.
\label{sum_gal}
\end{equation}
The first term on the right hand side of Eq. (\ref{sum_gal}) is the result of
a coherent sum of terms corresponding to mode $\vec{k}$.
The second term is the result of
an incoherent sum which we have approximated to
$N_{\textrm{gal}}^{1/2} e^{i \phi}$, where the phase $\phi$ is arbitrary. 
We define the ``galaxy power spectrum"
\begin{equation}
P_{\textrm{gal}}(\vec{k}) \equiv V \left| \Delta_{\vec{k}} \right|^2,
\label{P_gal0}
\end{equation}
and obtain
\begin{eqnarray}
P_{\textrm{gal}}(\vec{k}) & = & V \left| \frac{1}{N_{\textrm{gal}}}
  \sum_j e^{-i\vec{k} \cdot \vec{X_j}} \right|^2 
  - \frac{V}{N_{\textrm{gal}}} \nonumber \\
& & \pm \sqrt{\frac{2 V}{N_{\textrm{gal}}} P_{\textrm{gal}}(k)}.
\label{P_gal}
\end{eqnarray}
The transition between signal and noise occurs at
$\log_{10}(P_{\textrm{gal}}(k) / h^{-3} \textrm{ Mpc}^3) 
\approx 3.47$ for our data sample,
and $\approx 3.49$ for our reference simulation. To test these ideas
we can select a narrow range of MAGr, MAGg, or MAG to
shift the noise upwards, compare Figs. \ref{Pgal_k_SDSS_DR14}, 
\ref{P_gal_k_SDSS14_2}, and \ref{P_gal_k_SDSS14_2_g}
(which plot the first term on the 
right hand side of Eq. (\ref{P_gal}) and include
the noise at large $k$).

Averaging over $\vec{k}$ in a bin of $k \equiv | \vec{k} |$ 
obtains $P_{\textrm{gal}}(k)$.
The factor $V$ is 
inserted so that $P_{\textrm{gal}}(k)$ becomes 
independent of the arbitrary choice of $V$ for large $V$.
The function $P_{\textrm{gal}}(k)$ defines statistically
the distribution of galaxies.
The variables $\vec{k}$ in Eqs. (\ref{delta_c}) and (\ref{P_gal}) should not
be confused: there is not necessarily a 
one-to-one relation between them.

Results for data are presented in
Figs. \ref{Pgal_k_SDSS_DR14}, \ref{P_gal_k_SDSS14_2}, 
and \ref{P_gal_k_SDSS14_2_g}.
We note that the galaxy bias $b$
depends on MAGr and MAGg.
Even tho $N_{\textrm{gal}} \gg N_{\vec{k}}$,
$N_{\vec{k}} > N_{\textrm{gal}}^{1/2}$ at small $k$.
For this reason $P_{\textrm{gal}}(k)$ in 
Fig. \ref{Pgal_k_SDSS_DR14} extends to higher $k$ than
in Figs. \ref{P_gal_k_SDSS14_2} and \ref{P_gal_k_SDSS14_2_g} 
before saturating with noise. 
Figure \ref{Pgal_k_SDSS_DR14_less_noise} presents
the noise subtracted galaxy power spectrum
$P_\textrm{gal}(k)$, obtained from
Fig. \ref{Pgal_k_SDSS_DR14}, compared with $P(k)$
calculated with the indicated parameters.
Their ratio is the bias $b^2$.

Results for the simulations are presented in
Figs. \ref{P_gal_k_MC_2_run29_run30_run31},
\ref{P_gal_k_MC_2_run30_run32_run34}, and
\ref{P_gal_k_MC_2_run29_run52_run53}.
In Fig. \ref{P_gal_k_MC_2_run29_run52_run53} we compare
the reference simulation with $P'(k)$, with simulations
with $P'(k) \cdot (-\log_{10}(k/h \textrm{ Mpc}^{-1}))$
(``steeper slope"), or 
$P'(k) / (-\log_{10}(k/h \textrm{ Mpc}^{-1}))$
(``less slope"). Note that the function
$-\log_{10}(k/h \textrm{ Mpc}^{-1})$ varies between
$\approx 1.3$ to $\approx 0.5$ in the region of interest.
We observe, qualitatively, that the slope 
of $P(k)$ has a larger effect
on $P_\textrm{gal}(k)$ than the amplitude $A$.
A comparison of	the simulations	in Fig.	\ref{P_gal_k_MC_2_run29_run52_run53} with
$P_\textrm{gal}(k)$ from data in Fig. \ref{Pgal_k_SDSS_DR14} favors
a power	spectrum $P(k)$	``steeper" than	in the reference
simulation. 
The reference simulation has parameters	of $P(k)$ similar to the ones
obtained from the fit in Fig. \ref{P_vs_k} which assumes scale invariant
$b$, and $\sum m_\nu = 0$ eV.
The reference simulation is also similar to the fit ``$\sum m_\nu = 0.014$ eV"
in Fig. \ref{Pgal_k_SDSS_DR14_less_noise}
(taken from Fig. \ref{Pk_graph_log_2_wide_Weinberg} 
which assumes scale invariant $b$).
A steeper $P(k)$ implies $\sum m_\nu > 0$ as shown
in Fig.	\ref{Pgal_k_SDSS_DR14_less_noise} by the curve ``$\sum m_\nu = 0.719$ eV",
and corresponds	to a bias $b$ with positive slope as in	
Eq. (\ref{summnu7}) below.

\section{Luminosity and mass distributions of galaxies}

\begin{figure}
\includegraphics[scale=0.47]{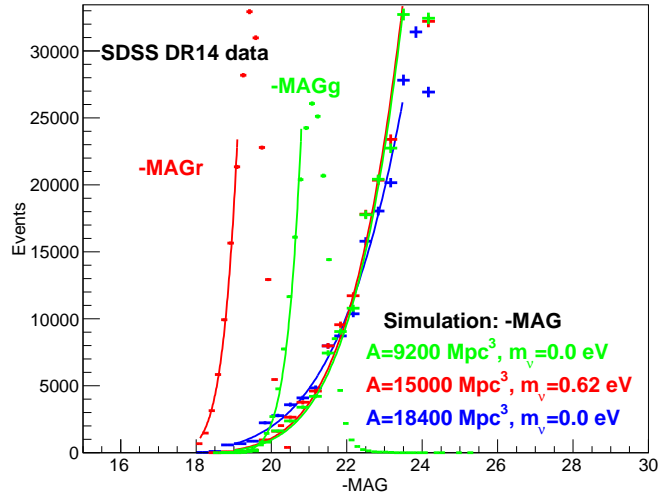}
\caption{\label{lum_vs_mass_29_30_31}
Distributions of MAGr and MAGg of SDSS DR14 
data, and distributions of MAG of several simulations
(see definitions in Section \ref{Data}).
The difference between the
MAGr or MAGg of data and MAG of simulations determines 
the ``mean" galaxy $L/M$ ratio.
}
\end{figure}

\begin{figure}
\includegraphics[scale=0.47]{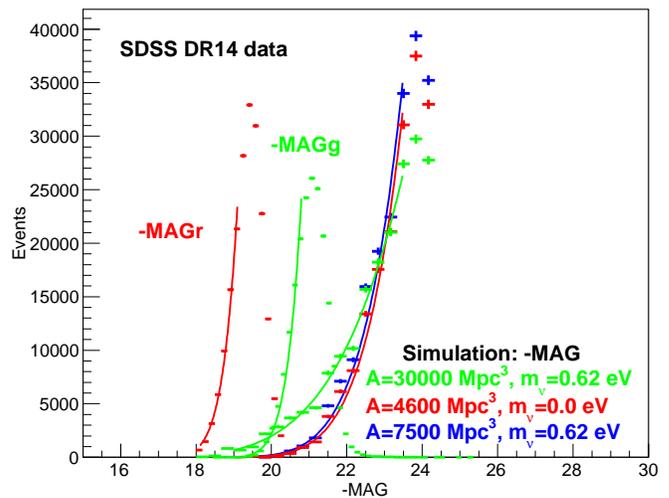}
\caption{\label{lum_vs_mass_32_33_34}
Same as Fig. \ref{lum_vs_mass_29_30_31} with additional simulations.
}
\end{figure}

Distributions of MAGr and MAGg from data, and MAG from 
several simulations are presented in Figs. 
\ref{lum_vs_mass_29_30_31} and \ref{lum_vs_mass_32_33_34}.
From these figures it is possible to obtain the
``mean" luminosity-to-mass ratios. We note that these figures
do not show useful sensitivity to $\sum m_\nu$.

\section{Test of scale invariance of the
galaxy bias $b$}

\begin{table*}
\caption{\label{sphere_counts}
Mean galaxy counts $\bar{N}$ in spheres of radius $r_s$.
All spheres have their center at redshift $z = 0.5$.
The number of spheres is $N_s = N_y \times N_z$.
Note that the observed root-mean-square (rms) fluctuation 
relative to the mean $\bar{N}$ is larger
than the corresponding statistical fluctuation, i.e.
rms$/\bar{N} > 1/\sqrt{\bar{N}}$.
$\sigma_{r_s/h}$ is calculated with Eqs. (\ref{Pk})
and (\ref{s8}) with $N^2$
chosen so $\sigma_8 = 0.770$ to set the scale for $b$
(e.g. $N^2 = 0.8472 \times 10^{-10}$
for $\sum m_\nu =0$ eV, or
$N^2 = 1.3575 \times 10^{-10}$
for $\sum m_\nu =0.6$ eV).
Both the galaxy counts and $\sigma_{r_s/h}$ are 
obtained with the top-hat window function.
The true standard deviation is obtained from
$\sigma^2 = \textrm{rms}^2 -\bar{N} \pm 2 \sigma \sqrt{\bar{N}/N_s}$.
The measured ``bias" is defined as 
$b \equiv (\sigma/\bar{N}) / (0.779 \cdot \sigma_{r_s/h})$.
The last column is the $\chi^2$ of the five $b$'s
of spheres with $\bar{N} > 1$, assuming these $b$'s are
scale invariant. 
$h = 0.678$, $n = 1.0$, and $\Omega_m = 0.281$.
}
\begin{ruledtabular}
\begin{tabular}{l|ccccccr}
$r_s/h$ [Mpc] & 8     & 16    & 32    & 64    & 128    & 256 &  \\
$r_s$ [Mpc]   & 11.80 & 23.60 & 47.20 & 94.40 & 188.79 & 377.58 & \\
$N_y \times N_z$ & $151 \times 55$ & $75 \times 27$ & $37 \times 13$ 
  & $19 \times 7$ & $9 \times 3$ & $4 \times 1$ & \\ 
$\bar{N}$        & 0.836 & 6.781 & 52.279 & 410.74 & 3092.3 & 21810.0 & \\
$1/\sqrt{\bar{N}}$ & 1.0935 & 0.3840 & 0.1383 & 0.0493 & 0.0180 & 0.0068 \\
rms$/\bar{N}$      & 1.798 & 0.873 & 0.443 & 0.210 & 0.0870 & 0.0346 & \\
$\sigma / \bar{N}$ & $1.427 \pm 0.012$ & $0.784 \pm 0.009$ & 
  $0.421 \pm 0.006$ & $0.204 \pm 0.004$ & $0.085 \pm 0.003$ & $0.034 \pm 0.003$ & \\
$\sigma_{r_s/h}, \sum m_\nu = 0.0$ eV & 0.7700 & 0.4457 & 0.2255 & 0.0987 
  & 0.0374 & 0.0124 & $\chi^2$ \\
$b$, $\sum m_\nu = 0.0$ eV & $2.380 \pm 0.020$ & $2.257 \pm 0.025$ 
  & $2.398 \pm 0.036$ & $2.650 \pm 0.056$ 
  & $2.925 \pm 0.119$ & $3.503 \pm 0.349$ & 79.2 \\
$\sigma_{r_s/h}, \sum m_\nu = 0.3$ eV & 0.7700 & 0.4514 & 0.2321 & 0.1036 
  & 0.0402 & 0.0136 & \\
$b$, $\sum m_\nu = 0.3$ eV & $2.380 \pm 0.020$ & $2.228 \pm 0.024$ 
  & $2.329 \pm 0.035$ & $2.523 \pm 0.053$ 
  & $2.722 \pm 0.111$ & $3.193 \pm 0.318$ & 49.5 \\
$\sigma_{r_s/h}, \sum m_\nu = 0.6$ eV & 0.7700 & 0.4603 & 0.2425 & 0.1113 
  & 0.0443 & 0.0152 & \\
$b$, $\sum m_\nu = 0.6$ eV & $2.380 \pm 0.020$ & $2.185 \pm 0.024$ 
  & $2.230 \pm 0.033$ & $2.350 \pm 0.049$ 
  & $2.468 \pm 0.100$ & $2.862 \pm 0.285$ & 20.0 \\
$\sigma_{r_s/h}, \sum m_\nu = 0.7$ eV & 0.7700 & 0.4640 & 0.2468 & 0.1144 
  & 0.0460 & 0.0158 & \\
$b$, $\sum m_\nu = 0.7$ eV & $2.380 \pm 0.020$ & $2.168 \pm 0.024$ 
  & $2.191 \pm 0.033$ & $2.285 \pm 0.048$ 
  & $2.379 \pm 0.097$ & $2.755 \pm 0.275$ & 12.6 \\
$\sigma_{r_s/h}, \sum m_\nu = 0.8$ eV & 0.7700 & 0.4682 & 0.2516 & 0.1179 
  & 0.0478 & 0.0165 & \\
$b$, $\sum m_\nu = 0.8$ eV & $2.380 \pm 0.020$ & $2.148 \pm 0.023$ 
  & $2.149 \pm 0.032$ & $2.218 \pm 0.047$ 
  & $2.289 \pm 0.093$ & $2.648 \pm 0.264$ & 7.1 \\
$\sigma_{r_s/h}, \sum m_\nu = 0.9$ eV & 0.7700 & 0.4729 & 0.2570 & 0.1218 
  & 0.0497 & 0.0171 & \\
$b$, $\sum m_\nu = 0.9$ eV & $2.380 \pm 0.020$ & $2.127 \pm 0.023$ 
  & $2.104 \pm 0.032$ & $2.147 \pm 0.045$ 
  & $2.198 \pm 0.089$ & $2.541 \pm 0.253$ & 4.0 \\
$\sigma_{r_s/h}, \sum m_\nu = 1.0$ eV & 0.7700 & 0.4782 & 0.2630 & 0.1261 
  & 0.0519 & 0.0179 & \\
$b$, $\sum m_\nu = 1.0$ eV & $2.380 \pm 0.020$ & $2.103 \pm 0.023$ 
  & $2.056 \pm 0.031$ & $2.073 \pm 0.044$ 
  & $2.107 \pm 0.086$ & $2.434 \pm 0.243$ & 3.8 \\
$\sigma_{r_s/h}, \sum m_\nu = 1.2$ eV & 0.7700 & 0.4911 & 0.2775 & 0.1363 
  & 0.0568 & 0.0196 & \\
$b$, $\sum m_\nu = 1.2$ eV & $2.380 \pm 0.020$ & $2.048 \pm 0.022$ 
  & $1.948 \pm 0.029$ & $1.919 \pm 0.040$ 
  & $1.923 \pm 0.078$ & $2.220 \pm 0.221$ & 13.7 \\
\end{tabular}
\end{ruledtabular}
\end{table*}

In this Section we test the scale invariance of the
bias $b$ defined in Eq. (\ref{Pgal_Pb}).
To do so, we count galaxies in an array of $N_s = N_x \times N_y$
spheres of radii $r_s$, and obtain their mean $\bar{N}$, 
and their root-mean-square
(rms). All spheres have their center at redshift $z = 0.5$ to
ensure the homogeneity of the galaxy selections.
The results for $r_s = 8/h, 16/h, 32/h, 64/h, 128/h$, 
and $256/h$ Mpc are presented in Table \ref{sphere_counts}.
The $($rms$)^2$ has a contribution $\sigma^2$ from $P(k)$, and a 
contribution $\bar{N}$ from statistical fluctuations:
\begin{equation}
\sigma^2 = \textrm{rms}^2 - \bar{N} \pm 2 \sigma \sqrt{\frac{\bar{N}}{N_s}}.
\end{equation}
We compare $\sigma / \bar{N}$ obtained from galaxy counts, with
the relative mass fluctuations $\sigma_{r_s/h}$ obtained from
Eqs. (\ref{Pk}) and (\ref{s8}). The ratio of these two 
quantities divided by a correction factor
$C(\Omega_\Lambda / (\Omega_m (1 + z)^3))
/(C(\Omega_\Lambda / \Omega_m) (1 + z)) = 0.779$
\cite{Weinberg} is the bias $b$.

\begin{figure}
\includegraphics[scale=0.47]{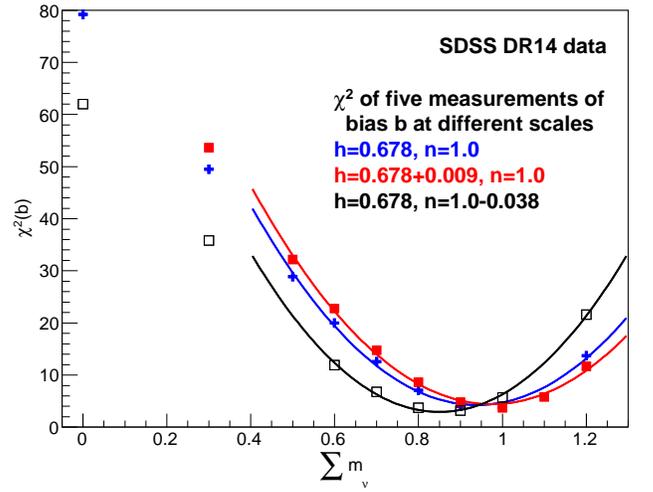}
\caption{\label{spheres_fit}
$\chi^2$ of five measurements of bias $b$
assumed to be scale invariant with respect
to their weighted mean
as a function of $\sum m_\nu$, for several
values of the Hubble parameter $h$, and the
spectral index $n$.
The five measurements of $b$ correspond to
scales $r_s = 16/h, 32/h, 64/h, 128/h$, and $256/h$ Mpc.
}
\end{figure}

The measured bias $b$ is a function of $r_s$, $\sum m_\nu$, $h$ and 
the spectral index $n$. Results for $h=0.678$ and $n=1$ are presented
in Table \ref{sphere_counts}. 
The last column is the $\chi^2$ of the five $b$'s
of spheres with $\bar{N} > 1$, assuming these $b$'s are
scale invariant with respect to their weighted
average.
Additional measurements of $\chi^2$ are presented 
in Fig. \ref{spheres_fit}. Assuming that $b$ is scale
invariant we obtain
\begin{equation}
\sum m_{\nu} = 0.939 + 0.035 \cdot \delta h + 0.089 \cdot \delta n \pm 0.008 \textrm{ eV},
\label{summnu_spheres}
\end{equation}
with minimum $\chi^2 = 3.2$ for four degrees of freedom.
We have defined $\delta h \equiv (h - 0.678)/0.009$, and
$\delta n \equiv (n - 1)/0.038$.

In conclusion, the galaxy bias $b$ is scale invariant within
the statistical uncertainties of $b$ presented in Table
\ref{sphere_counts}, provided $\sum m_{\nu}$ 
satisfies Eq. (\ref{summnu_spheres}),
else scale invariance is broken.
Note in Table \ref{sphere_counts} that the variation of $b$ with scale
depends on $\sum m_\nu$.

\section{Measurement of neutrino masses with
the Sachs-Wolfe effect and $\sigma_8$}
\label{SW_s8}

\begin{figure}
\includegraphics[scale=0.47]{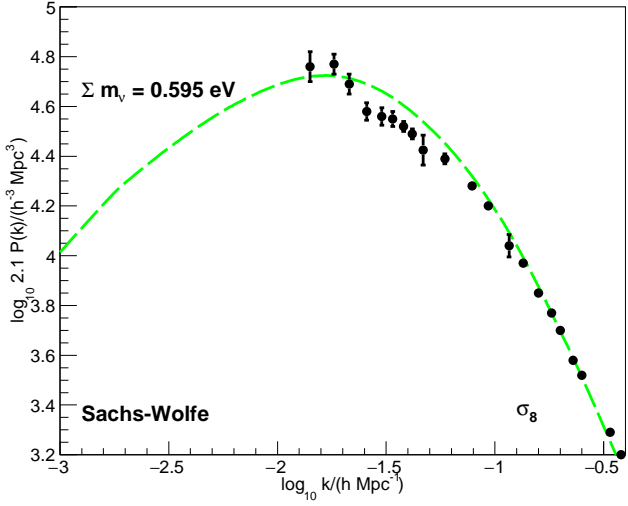}
\caption{\label{Pk_graph_Weinberg_fit_SW_s8}
Comparison of $P_\textrm{gal}(k)$ obtained
from the SDSS-III BOSS survey \cite{Pk_BOSS}
(``reconstructed") with $b^2 P(k)$
obtained from a
fit of Eq. (\ref{Pk})
to the Sachs-Wolfe effect and $\sigma_8$ only.
The fit obtains $\sum m_\nu = 0.595 \pm 0.225$ eV with
zero degrees of freedom. 
$h = 0.678$ and $n = 1.0$ are fixed.
}
\end{figure}

\begin{figure}
\includegraphics[scale=0.47]{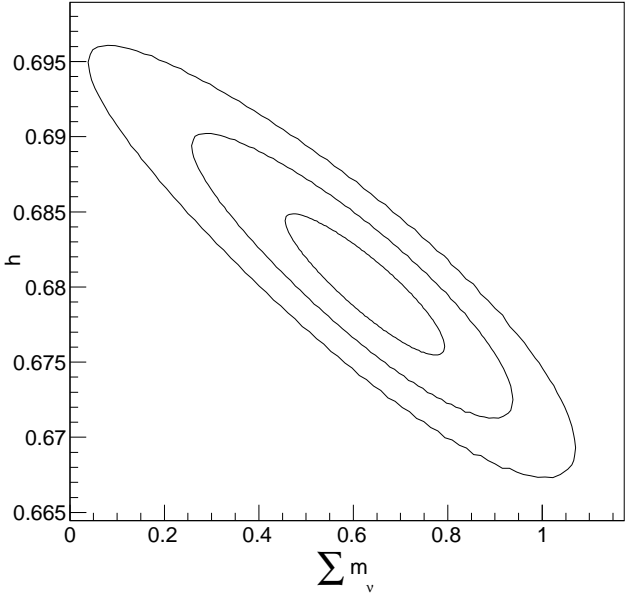}
\caption{\label{P_k_fit_Weinberg_h.C_bck080518}
Contours corresponding to
1, 2, and 3 standard deviations in the $(\sum m_\nu, h)$ plane,
from Sachs-Wolfe, $\sigma_8$, $h = 0.678 \pm 0.009$, and BAO measurements.
Points on the contours have $\chi^2 - \chi^2_\textrm{min} = 1, 4$, and 9,
respectively, where $\chi^2$ has been minimized	with respect
to $N^2$.
The total uncertainty of $\sum m_\nu$ is dominated by the
uncertainty of $h$.
In this figure $n = 1$, and the systematic uncertainties,
presented in Eq. (\ref{summnu_h}), are not included.
}
\end{figure}

\begin{figure}
\includegraphics[scale=0.47]{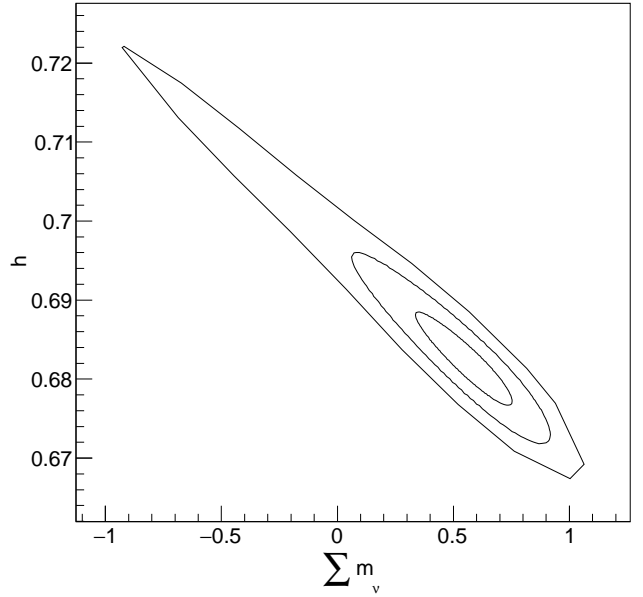}
\caption{\label{P_k_fit_Weinberg_h.C_bck230518_72}
Same as Fig. \ref{P_k_fit_Weinberg_h.C_bck080518} but
$h = 0.72 \pm 0.03$.
}
\end{figure}

We return to the measurement of neutrino masses.
Since the galaxy bias $b$ may be scale dependent,
in this Section we exclude measurements of
$P_\textrm{gal}(k)$ with galaxies.

The $\Lambda$CDM model is described by Eq. (\ref{Pk})
that has three free parameters: $N^2$, $n$, and $\sum m_\nu$.
We keep $n$ fixed.
We vary the two parameters $N^2$ and $\sum m_\nu$
to minimize a $\chi^2$
with two terms corresponding to two observables:
the Sachs-Wolfe effect
($N^2$ from Eq. (\ref{Nsq})),
and $\sigma_8$ given by Eq. (\ref{sigma_8}).
We therefore have zero degrees of freedom.
The result is a function of
$h$, $\Omega_m$, and the spectral index $n$, so we define
$\delta h \equiv (h - 0.678)/0.009$ \cite{PDG2016}, 
$\delta \Omega_m \equiv (\Omega_m - 0.281)/0.003$ \cite{BH_ijaa}, and
$\delta n \equiv (n - 1)/0.038$ \cite{PDG2016},
and obtain
\begin{eqnarray}
\sum m_\nu & = & 0.595 + 0.047 \cdot \delta h 
+ 0.226 \cdot \delta n + 0.022 \cdot \delta \Omega_m \nonumber \\
 & & \pm 0.225 \textrm{ (stat)} ^{+0.484}_{-0.152} \textrm{ (syst)}
\textrm{ eV}.
\label{summnu}
\end{eqnarray}
Note that in the ``6 parameter $\Lambda$CDM fit" \cite{PDG2016},
which assumes $\sum m_\nu = 0.06$ eV, $n = 0.968 \pm 0.006$.
Here, and below,
the systematic uncertainties are obtained by repeating the
fits with the top-hat window function instead of the
gaussian window function for $\sigma_8$ (and for $\sigma/\bar{N}$
if applicable), and also 
with $\sigma_8 = 0.815 \pm 0.009$ obtained with 
the ``6 parameter $\Lambda$CDM fit" \cite{PDG2016},
instead of $\sigma_8$ from direct
measurements, Eq. (\ref{sigma_8}).

The fit of Eq. (\ref{summnu}) is compared 
with measurements of $P_\textrm{gal}(k)$ obtained from
the SDSS-III BOSS survey \cite{Pk_BOSS} in 
Fig. \ref{Pk_graph_Weinberg_fit_SW_s8}.
It is interesting to note that the discrepancy, i.e. the drop of
$P(k)$ in the range $-1.6 < \log_{10}(k/h \textrm{ Mpc}) < -1.3$,
is also observed in Fig. \ref{Pgal_k_SDSS_DR14_less_noise}.

For comparison, reference \cite{BH_ijaa_2} obtains
\begin{equation}
\sum m_\nu = 0.711 - 0.335 \cdot \delta h
+ 0.050 \cdot \delta b \pm 0.063 \textrm{ eV},
\label{ijaa}
\end{equation}
where $\delta b \equiv (\Omega_b h^2 -0.02226)/0.00023$,
from a study of BAO with SDSS DR13 galaxies.
We allow $\Omega_b h^2$ to vary by one standard
deviation, i.e. $\delta b = 0 \pm 1$ \cite{BH_ijaa}.
To combine the independent measurements (\ref{summnu}) and (\ref{ijaa})
we add one more term to the $\chi^2$ corresponding
to the measurement $(\ref{ijaa})$, 
so we now have one degree of freedom. We obtain
\begin{eqnarray}
\sum m_\nu & = & 0.696 - 0.281 \cdot \delta h 
+ 0.032 \cdot \delta n + 0.003 \cdot \delta \Omega_m \nonumber \\
 & & \pm 0.075 \textrm{ (stat)} ^{+0.055}_{-0.029} \textrm{ (syst)}
\textrm{ eV},
\label{summnu2}
\end{eqnarray}
with $\chi^2 = 0.25$ for one degree of freedom, 
so the two independent measurements
of $\sum m_\nu$, Eqs. (\ref{summnu}) and (\ref{ijaa}), are consistent.
Note that the uncertainty of $h$ dominates the uncertainty of $\sum m_\nu$
in Eq. (\ref{summnu2}).

We now free $h$ and add one term to the $\chi^2$
corresponding to $h = 0.0678 \pm 0.009$ \cite{PDG2016}, and obtain
\begin{eqnarray}
\sum m_\nu & = & 0.633 \pm 0.168 \textrm{ (stat)} 
^{+0.064}_{-0.043} \textrm{ (syst) eV}, 
\label{summnu_h} \\
h & = & 0.680 \pm 0.005 \textrm{ (stat)},
\end{eqnarray}
with $\chi^2 = 0.07$ for one degree of freedom.
The systematic uncertainties in Eq. (\ref{summnu_h})
now include $\delta n$.
The $1 \sigma$, $2 \sigma$, and $3 \sigma$ contours are presented in
Fig. \ref{P_k_fit_Weinberg_h.C_bck080518}.

If instead we set
$h = 0.72 \pm 0.03$ from the direct measurement 
of the Hubble expansion rate \cite{PDG2016}, we obtain
\begin{eqnarray}
\sum m_\nu & = & 0.563 \pm 0.207 \textrm{ (stat)}
^{+0.064}_{-0.043} \textrm{ (syst) eV}, 
\label{summnu_h_73} \\
h & = & 0.682 \pm 0.006 \textrm{ (stat)},
\end{eqnarray}
with $\chi^2 = 1.7$ for 1 degree of freedom.
The corresponding $1 \sigma$, $2 \sigma$, and $3 \sigma$ 
contours are presented in 
Fig. \ref{P_k_fit_Weinberg_h.C_bck230518_72}.

\section{Measurement of neutrino masses with
the Sachs-Wolfe effect, $\sigma_8$ and $P_\textrm{gal}(k)$}

\begin{figure}
\includegraphics[scale=0.47]{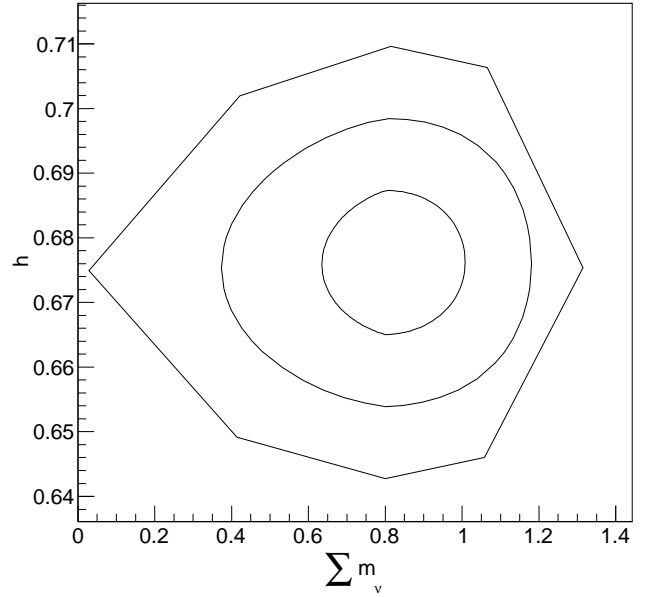}
\caption{\label{P_k_fit_Weinberg_h_b_BOSS.C_bck310518_n}
Contours corresponding to
1, 2, and 3 standard deviations in the $(\sum m_\nu, h)$ plane,
from Sachs-Wolfe, $\sigma_8$, $h = 0.678 \pm 0.009$, and 
$P_\textrm{gal}(k)$ measurements.
Points on the contours have $\chi^2 - \chi^2_\textrm{min} = 1, 4$, and 9,
respectively, where $\chi^2$ has been minimized	with respect
to $N^2$, $n$, $b_0$, and $b_1$.
}
\end{figure}

We repeat the fit of Fig. \ref{Pk_graph_log_2_wide_Weinberg}, 
which includes the ``reconstructed"
SDSS-III BOSS $P_\textrm{gal}(k)$ measurements \cite{Pk_BOSS}, but this time
we allow the galaxy bias $b$ to depend on scale:
$b \equiv b_0 + b_1 \log_{10}(k / h \textrm{ Mpc}^{-1})$.
Minimizing the $\chi^2$ with respect to
$\sum m_\nu$, $N^2$, $n$, $h = 0.678 \pm 0.009$, $b_0$,
and $b_1$, we obtain
\begin{eqnarray}
\sum m_\nu & = & 0.80 \pm 0.23 \textrm{ eV}, \nonumber \\
N^2 & = & (1.88 \pm 0.39) \times 10^{-10}, \nonumber \\ 
n & = & 1.064 \pm 0.068, \nonumber \\
h & = & 0.676 \pm 0.011, \nonumber \\ 
b_0 & = & 2.35 \pm 0.36, \nonumber \\ 
b_1 & = & 0.229 \pm 0.094, 
\label{summnu7}
\end{eqnarray}
with $\chi^2 = 27.8$ for $18$ degrees of freedom.
The uncertainties have been multiplied by $\sqrt{(27.8/18)}$.
Confidence contours are presented in 
Fig. \ref{P_k_fit_Weinberg_h_b_BOSS.C_bck310518_n}.
Fixing $b_1 = 0$ obtains $\chi^2 = 36.3$, so including
the scale dependence of $b$ is necessary.

\section{Measurement of neutrino masses with the Sachs-Wolfe
effect, $\sigma_8$, and galaxy fluctuations}

We repeat the measurements of $\sum m_\nu$ of
Section \ref{SW_s8} but add 4 more experimental
constraints: $\sigma /\bar{N}$ of galaxy counts in spheres of radius
$r_s = 16/h, 32/h, 64/h$, and $128/h$ Mpc, which are listed in
Table \ref{sphere_counts}. Spheres of radius $8/h$ Mpc were
not considered because they have $\bar{N} < 1$. Spheres of radius
$256/h$ Mpc were excluded because there are only 4 spheres
of this radius, and the difference between the rms for
the top-hat and gaussian window functions turns out to be large
(while consistent results are obtained for the other radii).
We add two more parameters to be fit: $b_0$ and $b_s$ which
define the bias $b = b_0 - i_s b_s$, with
$i_s = 0, 1, 2, 3$ for $r_s = 16/h, 32/h, 64/h$, and $128/h$ Mpc,
respectively.
Note that we do not obtain a good fit with fixed bias $b = b_0$,
and so have introduced a ``bias slope" $b_s$.

From the Sachs-Wolfe effect, $\sigma_8$, and the 4
$\sigma /\bar{N}$ measurements we obtain
\begin{eqnarray}
\sum m_\nu & = & 0.618 + 0.042 \cdot \delta h 
+ 0.206 \cdot \delta n + 0.019 \cdot \delta \Omega_m \nonumber \\
 & & \pm 0.209 \textrm{ (stat)} ^{+0.420}_{-0.139} \textrm{ (syst)}
\textrm{ eV},
\label{summnu3}
\end{eqnarray}
with $\chi^2 = 1.10$ for 2 degrees of freedom.
The variables that minimize the $\chi^2$ are
$\sum m_\nu$, $N^2$, $b_0$, and $b_s$.
This result may be compared with (\ref{summnu}).

Freeing $h = 0.678 \pm 0.009$, and keeping $n = 1.0$ fixed, we obtain
\begin{eqnarray}
\sum m_\nu & = & 0.618 \pm 0.214 \textrm{ eV}, \nonumber \\
N^2 & = & (2.11 \pm 0.31) \times 10^{-10}, \nonumber \\
h & = & 0.678 \pm 0.009, \nonumber \\
b_0 & = & 1.756 \pm 0.057, \nonumber \\
b_s & = & -0.062 \pm 0.042,
\label{summnu3_h}
\end{eqnarray}
with $\chi^2 = 1.10$ for 2 degrees of freedom.

Combining with the BAO measurement (\ref{ijaa}) we obtain
\begin{eqnarray}
\sum m_\nu & = & 0.697 - 0.276 \cdot \delta h 
+ 0.032 \cdot \delta n + 0.003 \cdot \delta \Omega_m \nonumber \\
 & & \pm 0.075 \textrm{ (stat)} ^{+0.055}_{-0.028} \textrm{ (syst)}
\textrm{ eV},
\label{summnu6}
\end{eqnarray}
with $\chi^2 = 1.29$ for 3 degrees of freedom.
The variables that minimize the $\chi^2$ are $\sum m_\nu$,
$N^2$, $b_0$, and $b_s$.
Freeing $h = 0.678 \pm 0.009$, and keeping $n = 1.0$ fixed, we obtain
\begin{eqnarray}
\sum m_\nu & = & 0.644 \pm 0.162 \textrm{ eV}, \nonumber \\
N^2 & = & (2.13 \pm 0.28) \times 10^{-10}, \nonumber \\
h & = & 0.680 \pm 0.005, \nonumber \\
b_0 & = & 1.756 \pm 0.057, \nonumber \\
b_s & = & -0.058 \pm 0.036,
\label{summnu4}
\end{eqnarray}
with $\chi^2 = 1.14$ for 3 degrees of freedom.

\begin{figure}
\includegraphics[scale=0.47]{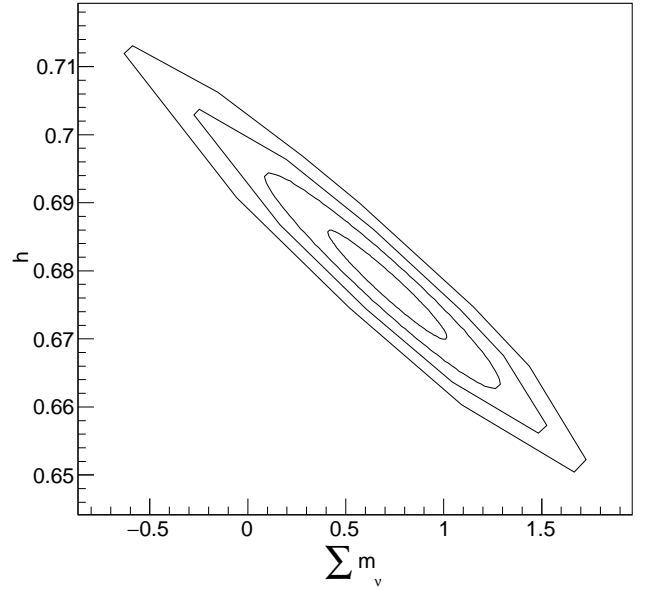}
\caption{\label{P_k_fit_Weinberg_h_b.C_bck270518}
Contours corresponding to 
1, 2, 3, and 4 standard deviations in the $(\sum m_\nu, h)$ plane,
from Sachs-Wolfe, $\sigma_8$, 4 $\sigma / \bar{N}$, BAO, 
and $h = 0.678 \pm 0.009$ measurements.
Points on the contours have $\chi^2 - \chi^2_\textrm{min} = 1, 4, 9$, and 16,
respectively, where $\chi^2$ has been minimized	with respect
to $N^2$, $n$, $b_0$, and $b_s$.
The total uncertainty of $\sum m_\nu$ is dominated by the
uncertainty of $h$.
In this figure the systematic uncertainties,
presented in Eqs. (\ref{summnu5}), are not included.
}
\end{figure}

Finally, freeing $n$, and minimizing the $\chi^2$ with respect to
$\sum m_\nu$, $N^2$, $n$, $h=0.678 \pm 0.009$, $b_0$, and $b_s$, 
we obtain
\begin{eqnarray}
\sum m_\nu & = & 0.719 \pm 0.312 \textrm{ (stat)}^{+0.055}_{-0.028} 
  \textrm{ (syst)} \textrm{ eV}, \nonumber \\
N^2 & = & (2.09 \pm 0.33) \times 10^{-10}, \nonumber \\
n & = & 1.021 \pm 0.075, \nonumber \\
h & = & 0.678 \pm 0.008, \nonumber \\
b_0 & = & 1.751 \pm 0.060, \nonumber \\
b_s & = & -0.053 \pm 0.041,
\label{summnu5}
\end{eqnarray}
with $\chi^2 = 1.1$ for 2 degrees of freedom.
The parameter correlation coefficients, 
defined in \cite{PDG2016}, are \\

\begin{tabular}{lcccccc}
 & $\sum m_\nu$ & $N^2$ & $n$              & $h$             & $b_0$             & $b_s$ \\
$\sum m_\nu$ & \phantom{-}1.000  & -0.019 & 0.856      & -0.966          & -0.226 & \phantom{-}0.779 \\
$N^2$ & -0.019            & \phantom{-}1.000    & -0.491 & 0.018  & -0.155            & \phantom{-}0.428 \\
$n$ & \phantom{-}0.856  & -0.491            & \phantom{-}1.000  & -0.834 & -0.303 & \phantom{-}0.427 \\
$h$ & -0.966            & \phantom{-}0.018    & -0.834 & 1.000  & \phantom{-}0.219  & -0.755 \\
$b_0$ & -0.226            & -0.155              & -0.303 & 0.219  & \phantom{-}1.000  & -0.037 \\
$b_s$ & \phantom{-}0.779 & \phantom{-}0.428 & \phantom{-}0.427 & -0.755  & -0.037  & \phantom{-}1.000 \\
\end{tabular}

Note that we have measured the amplitude
$N^2$ and spectral index $n$ of $P(k)$, and the
bias $b_0$ including its
slope $b_s$ for the SDSS DR14 galaxy selections
at redshift $z = 0.5$. 1, 2, 3, and 4 standard deviation
contours are presented in Fig. \ref{P_k_fit_Weinberg_h_b.C_bck270518}.

Figure \ref{P_k_fit_Weinberg_h_b.C_bck270518} and 
Eq. (\ref{summnu5}) are our final results.

\section{Conclusions}

We have studied galaxy distributions with 
Sloan Digital Sky Survey SDSS DR14 data
and with simulations searching for variables that can
constrain neutrino masses. 
Fitting the predictions of the
$\Lambda$CDM model to the Sachs-Wolfe effect, $\sigma_8$,
$P_\textrm{gal}(k)$, fluctuations of galaxy counts in
spheres	of radii ranging from $16/h$ to	$128/h$	Mpc,    
BAO measurements,
and $h = 0.678 \pm 0.009$, in various combinations, \textit{with
free spectral index $n$}, and free galaxy bias \textit{and 
galaxy bias slope}, we obtain consistent          
measurements of	$\sum m_\nu$. 
The uncertainty of $\sum m_\nu$ is dominated by the 
uncertainty of
$h$, so we have presented
confidence contours in the $(\sum m_\nu, h)$ plane. 

Fitting the predictions of the
$\Lambda$CDM model to the Sachs-Wolfe effect and $\sigma_8$
we obtain (\ref{summnu}).
Fitting the predictions of the
$\Lambda$CDM model to the Sachs-Wolfe effect, $\sigma_8$,
and galaxy number fluctuations $\sigma / \bar{N}$ in
spheres of radius $r_s = 16/h, 32/h, 64/h$, and $128/h$,
we obtain (\ref{summnu3}).
These results are consistent with the measurement (\ref{ijaa}) with BAO.
Combining these last two independent measurements we obtain
\begin{eqnarray}
\sum m_\nu & = & 0.697 - 0.276 \cdot \delta h
+ 0.032 \cdot \delta n + 0.003 \cdot \delta \Omega_m \nonumber \\
 & & \pm 0.075 \textrm{ (stat)} ^{+0.055}_{-0.028} \textrm{ (syst)}
\textrm{ eV}.
\end{eqnarray}
Note that the uncertainty of $\sum m_\nu$ is dominated
by the uncertainty of $h$. A global fit with
$h = 0.678 \pm 0.009$ obtains
$\sum m_\nu = 0.719 \pm 0.312 \textrm{ (stat)}^{+0.055}_{-0.028}
  \textrm{ (syst)}$ eV,
$h = 0.678 \pm 0.008$,
and the amplitude and spectral index of $P(k)$:
$N^2 = (2.09 \pm 0.33) \times 10^{-10}$, and
$n = 1.021 \pm 0.075$.
The fit also returns the galaxy bias $b$ including
its scale dependence.

Figure \ref{P_k_fit_Weinberg_h_b.C_bck270518} and
Eq. (\ref{summnu5}) are our final results.

\section{acknowledgement}

Funding for the Sloan Digital Sky Survey (SDSS) has been provided 
by the Alfred P. Sloan Foundation, the Participating Institutions, 
the National Aeronautics and Space Administration, the 
National Science Foundation, the U.S. Department of Energy, the 
Japanese Monbukagakusho, and the Max Planck Society. 
The SDSS Web site is http://www.sdss.org/.

The SDSS is managed by the Astrophysical Research Consortium (ARC) 
for the Participating Institutions. The Participating Institutions 
are The University of Chicago, Fermilab, the Institute for Advanced Study, 
the Japan Participation Group, The Johns Hopkins University, 
Los Alamos National Laboratory, the Max-Planck-Institute for Astronomy (MPIA), 
the Max-Planck-Institute for Astrophysics (MPA), New Mexico State University, 
University of Pittsburgh, Princeton University, 
the United States Naval Observatory, and the University of Washington.

\end{document}